\documentclass[reprint, superscriptaddress, amssymb, aps, pra, a4, showkeys]{revtex4-2}
\usepackage[utf8]{inputenc}
\usepackage{amsmath}
\usepackage{braket}
\usepackage{amsfonts}
\usepackage{amssymb}
\usepackage{xcolor}
\usepackage{graphicx}
\usepackage{titlesec}
\usepackage{hyperref}

\begin{document}

\title{Generating arbitrary non-separable states with polarization and orbital angular momentum of light}

\author{Sarika Mishra} \email{sarikamishra005@gmail.com}
\affiliation{Quantum Science and Technology Laboratory, Physical Research Laboratory, Ahmedabad, India 380009.}
\affiliation{Indian Institute of Technology, Gandhinagar, India 382355.}
\author{Ali Anwar}
\affiliation{Fraunhofer Centre for Applied Photonics, 99 George Street, Glasgow, Scotland, United Kingdom.}
\author{R. P. Singh} \email{rpsingh@prl.res.in}
\affiliation{Quantum Science and Technology Laboratory, Physical Research Laboratory, Ahmedabad, India 380009.}

\date{\today}

\begin{abstract}
\noindent We demonstrate an experimental method to generate arbitrary non-separable states of light using polarization and orbital angular momentum (OAM) degrees of freedom. We observe the intensity distribution corresponding to OAM modes of the light beam by projecting the non-separable state into different polarization states. We further verify the presence of non-separability by measuring the degree of polarization and linear entropy. This classical non-separability can be easily transferred to the quantum domain using spontaneous parametric down-conversion for applications in quantum communication and quantum sensing.
\end{abstract}

\keywords{Non-separable states of light, Vector vortex beam, Classical entanglement, Quantum entanglement, Orbital angular momentum.}

\maketitle

\section{\label{sec1}Introduction} 
\noindent In general, quantum entanglement \cite{horodecki2009quantum, terhal2002detecting} utilizes the property of quantum correlation between two particles of
similar nature or similar degree of freedom (DoF). However, entanglement can also be established between two different DoFs \cite{neves2009hybrid, nagali2010generation}, where one DoF of the system can not be measured without affecting the other. This shows the non-separable nature of entangled states, which means that one state of the system can not be separated from the other.\par
Importantly, non-separability can be explained in both domains, classical and in quantum \cite{shen2022nonseparable}. In quantum optics, entanglement between multiple photons with similar or different DoFs has been demonstrated, where the measurement of a DoF on one photon affects the states of others \cite{kwiat1995new, nagali2010generation,dada2011experimental, hamel2014direct}. A local entanglement between multiple intrinsic DoFs of a single photon \cite{chithrabhanu2016pancharatnam} is used to describe the non-separability in classical light that is controversially termed as “classical entanglement” \cite{spreeuw1998classical,ghose2014entanglement, karimi2015classical, paneru2020entanglement,luis2009coherence} due to its violation of Bell-like inequality \cite{borges2010bell, pires2010measurement}. Typical examples are spin-orbit vector beams, non-separable in the spatial and polarization DoFs \cite{pereira2014quantum, perumangatt2015scattering,salla2015recovering}. Some classes of vector beam such as Poincare beams and vector vortex beams \cite{galvez2012poincare, beckley2010full, chithrabhanu2016pancharatnam} show the non-separable nature between polarization and orbital angular momentum (OAM) of the beam. These classical non-separable states (CNSS) of light are mathematically
analogous to quantum entangled states \cite{aiello2015quantum}. This existing parallelism
between classical and quantum non-separable states has generated 
great interest for their applications in quantum communication \cite{ndagano2017characterizing, ndagano2018creation, milione2015using, krenn2015twisted, otte2018entanglement}, polarization metrology \cite{toppel2014classical}, quantum imaging \cite{magana2019quantum}, quantum computing gates \cite{daboul2003quantum} etc. \par

Employing more than one degree of freedom of light allows one to fetch more information through a single photon. Since OAM is related to the spatial mode and forms an infinite dimensional Hilbert space \cite{allen1992orbital, bai2022vortex}, this kind of multidimensionality offers a realization of d-dimensional qudits that increases the channel capacity in quantum communication \cite{willner2015optical,forbes2019quantum, rozenberg2023designing}. Due to the high channel capacity, the exploitation of high dimensional quantum state using of spatial modes has received a lot of recent interest in quantum communication field \cite{bacco2020quantum,cozzolino2019orbital, willner2015optical}. In recent years, a classical non-separable light using polarization and OAM (vector vortex beams) also attracted a lot of interest due to its direct transformation into a quantum hybrid entangled state \cite{jabir2017direct} which can be further utilized in quantum cryptography protocols \cite{krenn2015twisted, cozzolino2019air}. 

Generation and characterization of classical non-separable states of polarization and OAM have already been studied \cite{slussarenko2010polarizing,chithrabhanu2016pancharatnam,salla2015recovering, perumangatt2015scattering}. Various interferometric methods are used to generate vector vortex beams representing non-separable states of light with polarization and OAM, such as Michelson \cite{maurer2007tailoring}, folded Mach-Zehnder \cite{zhang2017efficient, liu2018highly}, beam displaced \cite{guo2021generation} and Saganac \cite{slussarenko2010polarizing} interferometers. Interferometric methods generally have inherent mode stability issues and require a fine-tuned alignment. Moreover, Sagnac interferometric method can only generate a non-separable state with a fixed OAM value with a spiral phase plate chosen. To overcome the limitations of interferometric methods, some all-in-line setups with q-plates \cite{chen2010single, cardano2012polarization} and spatial light modulators \cite{fu2016generating, mandal2020common} have been implemented. \par
In this article, we propose an  experimental method to generate an arbitrary non-separable state using spiral phase plate (SPP) \cite{beijersbergen1994helical} and spatial light modulator (SLM),  \cite{zhu2014arbitrary} which does not require any interferometric setup and fine-tuning of alignment. The proposed method is similar to Ref. \cite{fu2016generating, mandal2020common} where another SLM is used instead of the SPP. However, the scope of application of such non-separable states in the quantum domain is not explored in these works. Here, we quantify the non-separability of the states generated by this method, which clearly reflects the use of such a method in quantum experiments related to OAM \cite{jabir2017direct, krenn2015twisted}. To quantify the non-separability, we project the output state to different polarization and record the corresponding intensity distributions. Further, we verified the non-separability by measuring the degree of polarization and linear entropy \cite{gamel2012measures, de2014relationship, qian2011entanglement, perumangatt2015scattering}.\par
This article is organized as follows: In section \ref{sec2}, we develop the theory of our proposed idea. We first discuss the old method to generate classical non-separable state, and then we compare it with our proposed setup for arbitrary classical non-separable states of light using polarization and OAM. In section \ref{sec3} and \ref{sec3.0}, we explain all the details of our experiment and results demonstrating the creation and detection of these states. Finally, in section \ref{sec4}, we conclude the article indicating its potential applications.

\section{\label{sec2}Theory}
\begin{figure}[h!]
    \centering
    \includegraphics[width=8.6cm]{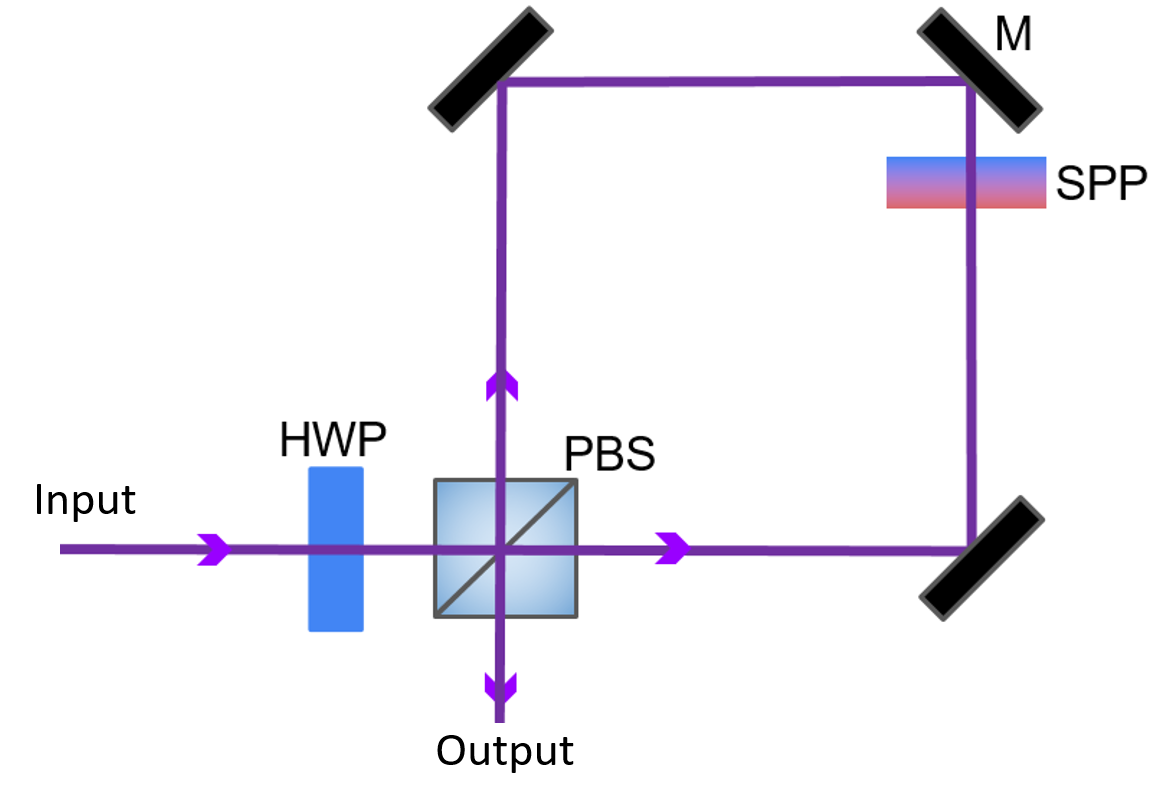}
    \caption{Sagnac interferometry based classical non-separable state}
    \label{ch7_NSS}
\end{figure}

\noindent Generally, a classical non-separable state in polarization and OAM DoF is generated using a polarizing Sagnac interferometer \cite{slussarenko2010polarizing, perumangatt2015scattering} as shown in Figure \ref{ch7_NSS}. 
In this experimental setup, a diagonally polarized Gaussian beam is fed into the Sagnac interferometer where a polarizing beam splitter (PBS) splits the horizontal and vertical polarization in two different directions, clockwise and counter-clockwise. Both the orthogonally polarized Gaussian beams counter propagate. An SPP of order $m$ converts the Gaussian beam to a vortex beam of order $m$ and $-m$ for horizontal and vertical polarized light respectively. Both the beams combine at the same PBS to form the non-separable state,
\begin{equation}
     \ket{\Psi}_{ns}=\frac{1}{\sqrt{2}}\left( \ket{H}\ket{m}+\ket{V}\ket{-m}\right).
    \label{ch7_4}
\end{equation}
 At the output of the Sagnac interferometer, a quarter-wave plate (QWP), a half-wave plate (HWP), and a PBS can be used for the measurement of the non-separable state. Projection to a particular polarization state gives the information about spatial mode (OAM) corresponding to that polarization.\par
The main drawback of the Sagnac interferometry based non-separable state is that it produces only a particular type of state as mentioned in Eqn \ref{ch7_4}, and it also requires a perfect alignment. To overcome this problem and to generate any arbitrary classical non-separable state, we propose an easier method.
\begin{figure}[h!]
    \centering
    \includegraphics[width=8cm]{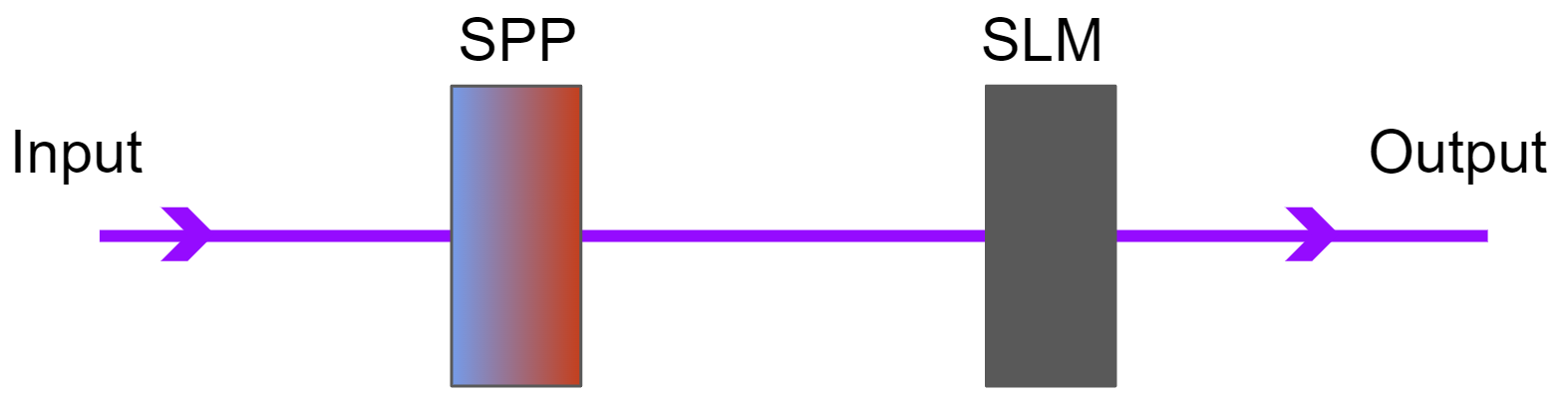}
    \caption{Proposed method to generate classical non-separable states. SPP: Spiral phase plate, SLM: Spatial light modulator}
    \label{NS}
\end{figure}
 In this method, if the input beam is a diagonally polarized Gaussian beam, then by using SPP and SLM in a row as shown in Figure \ref{NS} one can generate different non-separable states by varying the OAM values on the SLM. The main advantage of parallel aligned LC-SLM is that the orthogonal polarization state is unaffected by the SLM \cite{moreno2012complete}. In our case, the liquid crystal (LC) molecules inside the SLM are aligned in a horizontal direction, so it does not affect the vertically polarized light, which means that the spatial mode of vertically polarized beam will be unchanged after passing through the SLM. Since SLM is not adding any OAM value to vertical polarization, therefore in the final state both horizontal and vertical polarized beams will have different OAM values. The final state can be written as,
\begin{equation}
   \ket{\Psi}_{ns}=\frac{1}{\sqrt{2}}\left( \ket{H}\ket{m_{\text{SPP}}+m_{\text{SLM}}}+ \ket{V}\ket{m_{\text{SPP}}}\right). 
   \label{ch7_5}
\end{equation}
By changing the value $m_{\text{SLM}}$ via SLM one can generate the non-separable state of any order. This method is easy to use as compared to the Sagnac interferometry which requires a perfect alignment. Although the schematic shown in Figure \ref{NS} generates a class of NS states with a fixed OAM associated with the vertically polarized component of the beam, An SPP of different order can be used to change the OAM values associated with V polarized component. \par 
To generate a completely arbitrary NS state, the SPP in Figure \ref{NS} should be replaced with another SLM along with one HWP rotated at $45^{\circ}$. In this case, if the input beam is a diagonally polarized Gaussian beam then the combination of SLM and HWP (at $45^{\circ}$) can easily add any OAM value to V polarized component only, and the other SLM which is already shown in figure \ref{NS} will only change OAM of H polarized component. Hence, an arbitrary non-separable state can be generated using such a configuration.

\section{\label{sec3}Experimental setup}
\begin{figure}[h!]
    \centering
    \includegraphics[width=8.5cm]{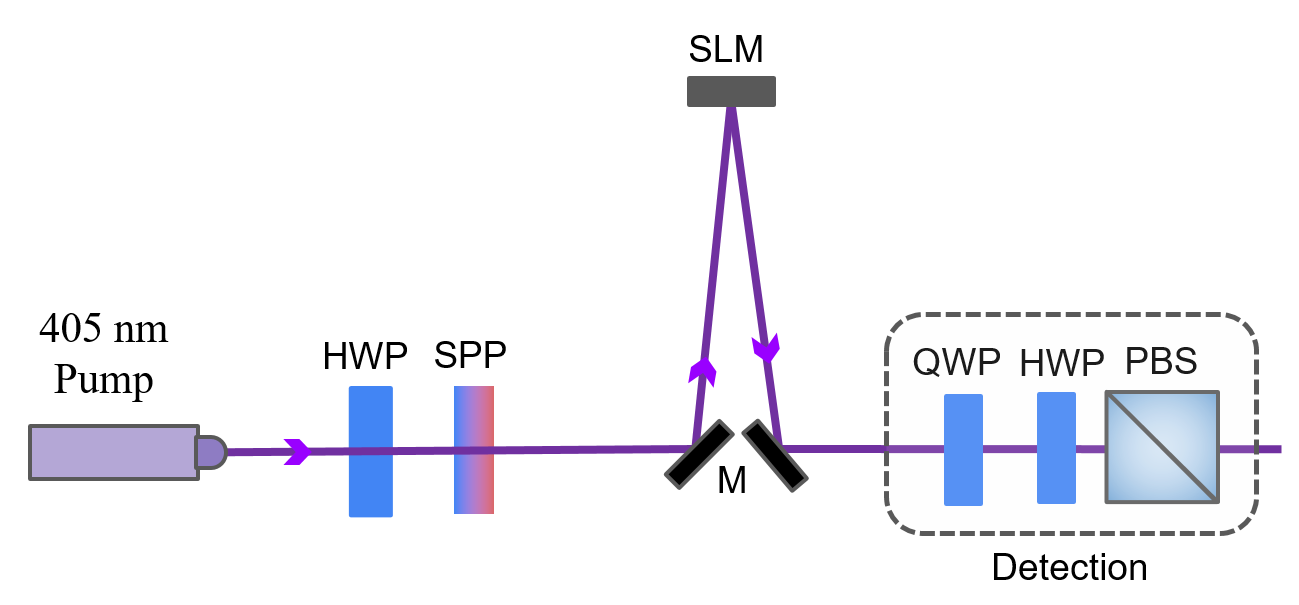}
    \caption{Experimental setup to generate classical non-separable state. SLM does not modulate the spatial mode of vertically polarized light. Therefore, at the output of SLM, $H$ and $V$ polarized light will have different OAM modes}
    \label{CH7_CNSS}
\end{figure}

 The experimental setup for the generation of a classical non-separable state is shown in Figure \ref{CH7_CNSS}. 
Toptica TopMode (405 nm) laser is used to perform the experiment. The horizontally polarized light with Gaussian mode ($\ket{H}\ket{0}$) is converted into the diagonal polarization ($\ket{D}\ket{0}$) after passing through the HWP which is oriented at 22.5$^{\circ}$ with respect to the fast axis. An SPP of order $m$ is kept in the path of the light beam. After passing through the SPP, the state of the light will be,
 \begin{align}
     \ket{\Psi} &= \ket{D}\ket{m_{\text{SPP}}} \nonumber \\
            &= \ket{H}\ket{m_{\text{SPP}}}+\ket{V}\ket{m_{\text{SPP}}},
            \label{ch7_6}
 \end{align}
The output of the SPP is then imaged onto the SLM. We imprinted a computer-generated hologram into the SLM to generate the vortex beam of order $m_{SLM}$. Since the liquid crystal molecules inside the SLM are aligned in a horizontal direction, it does not affect the vertical polarization. The SLM will only change the spatial profile of horizontally polarized light. The order of the spatial mode can be easily controlled by the SLM through a computer. The final state of the light beam after reflecting back from the SLM is written as,
 \begin{equation} \ket{\Psi}=\ket{H}\ket{m_{\text{SPP}}+m_{\text{SLM}}}+e^{i\phi}\ket{V}\ket{m_{\text{SPP}}}.
     \label{cc7_4}
 \end{equation}
 where $\phi$ is the relative phase delay between $H$ and $V$ polarized lights due to the birefringent walk-off inside the SLM.\par

 \section{\label{sec3.0}Results and discussion}
 
\begin{figure}[h!]
    \centering
    \includegraphics[width=8.6cm]{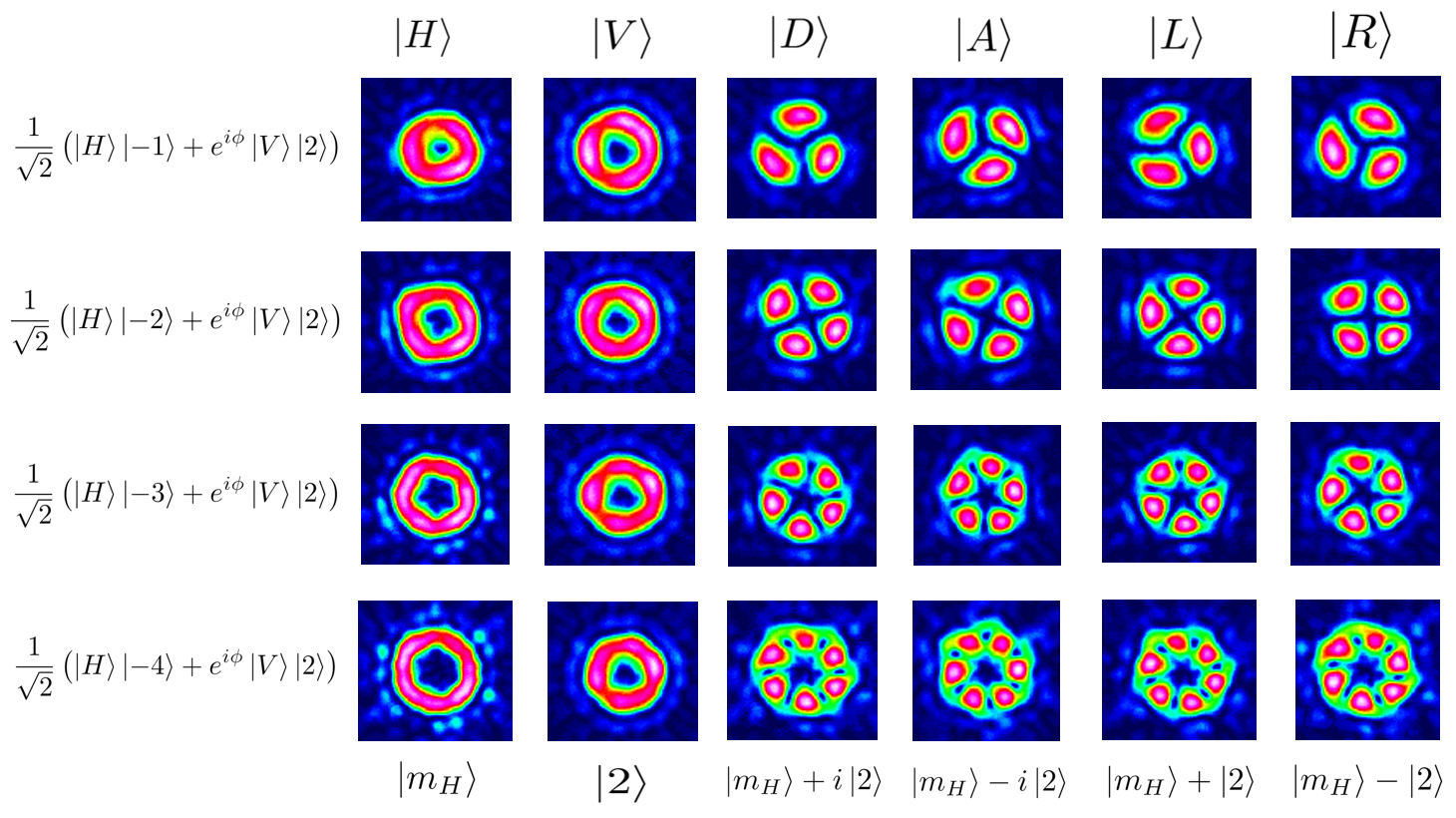}
    \caption{ Experimentally generated classical non-separable stats with different OAM. SPP of order $m=2$ is used. $m_H$ is the order of OAM associated with horizontally polarized light. $\phi$ is the relative phase between $H$ and $V$ polarized light.}
    \label{c7_11}
\end{figure}

\begin{figure}[h!]
    \centering
    \includegraphics[width=8.6cm]{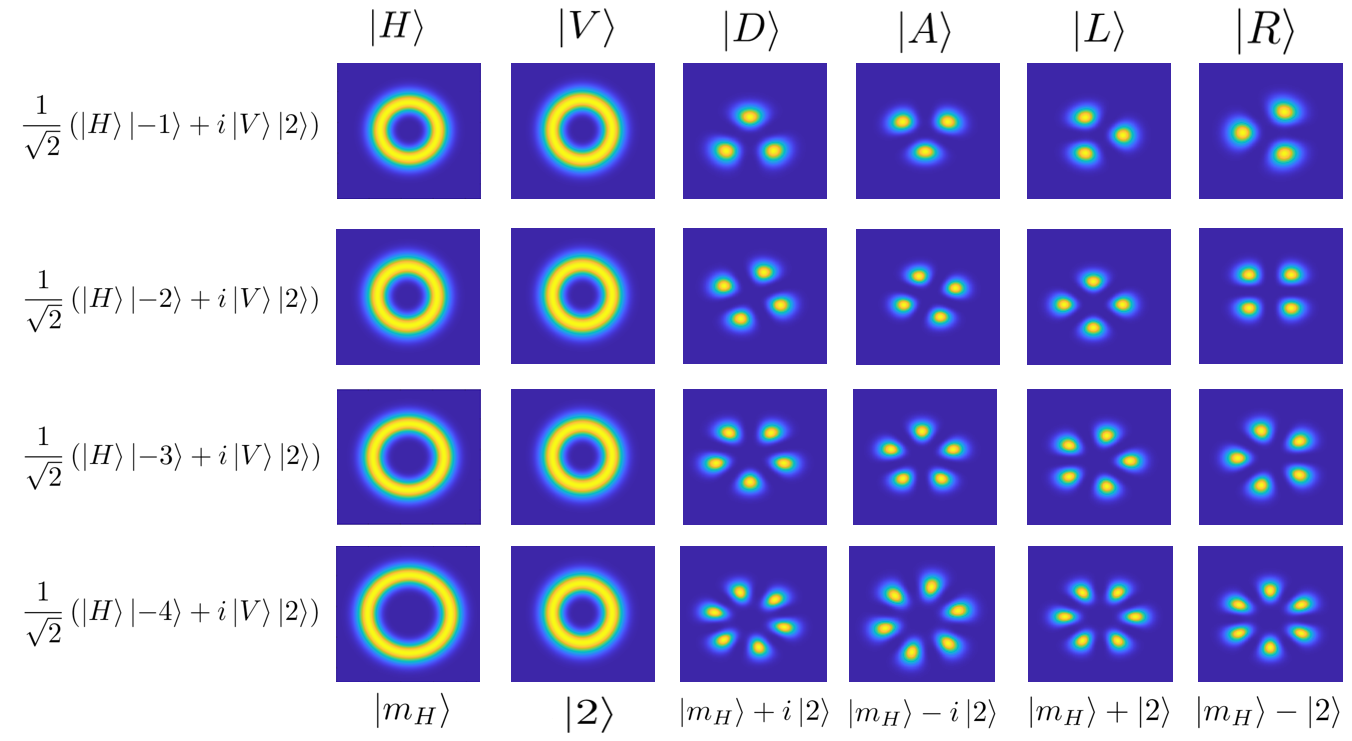}
    \caption{ Theoretically simulated intensity distributions of classical non-separable states for different polarization projections (linear (H/V and D/A) and circular (L/R)). SPP of order $m=2$ is used here and $m_H$ is the OAM value associated with horizontally polarized light.}
    \label{fig5}
\end{figure}

\begin{figure}[h!]
    \centering
    \includegraphics[width=8.5cm]{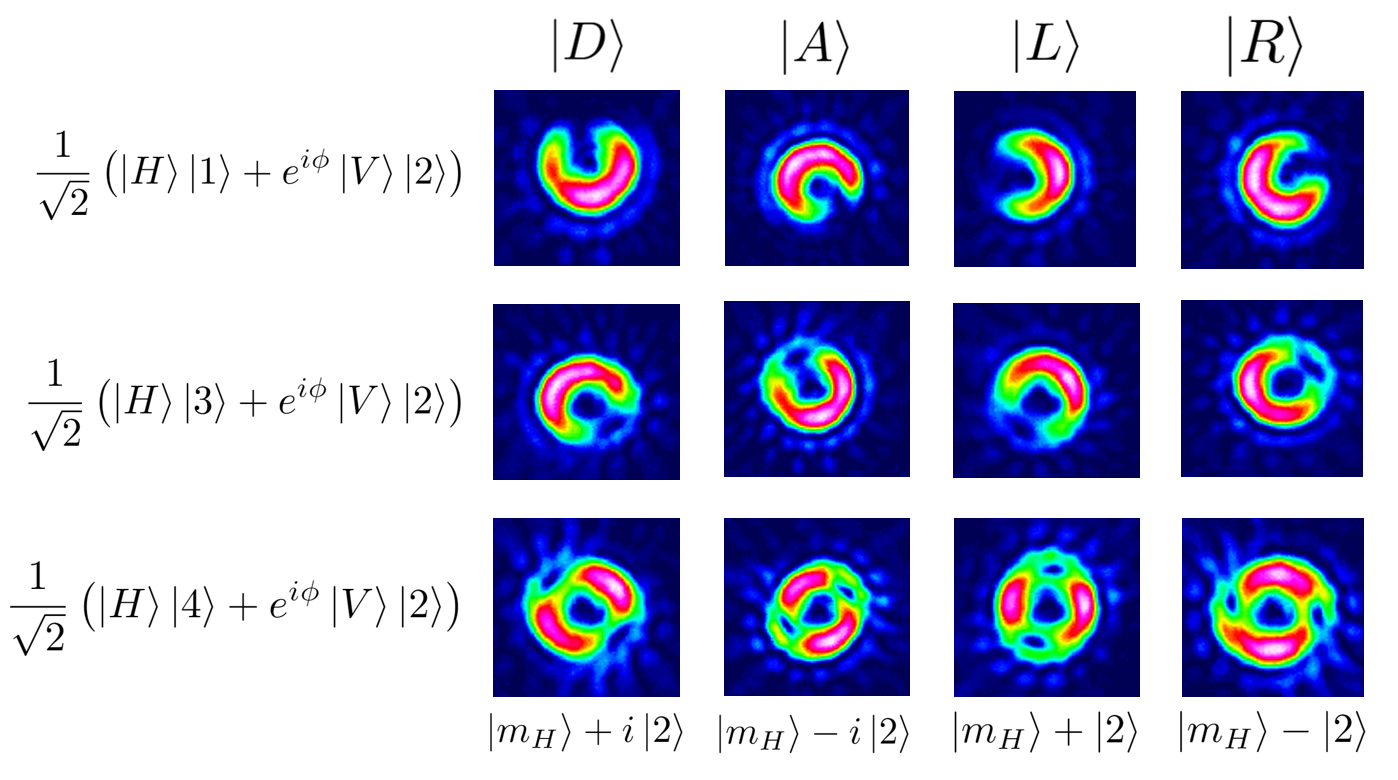}
    \caption{ Experimentally generated classical non-separable states with different OAM. SPP of order $m=2$ is used. $m_H$ is the order of OAM associated with horizontally polarized light. $\phi$ is the relative phase between $H$ and $V$ polarized light.}
    \label{exp:cnss}
\end{figure}

\begin{figure}[h!]
    \centering
    \includegraphics[width=8.5cm]{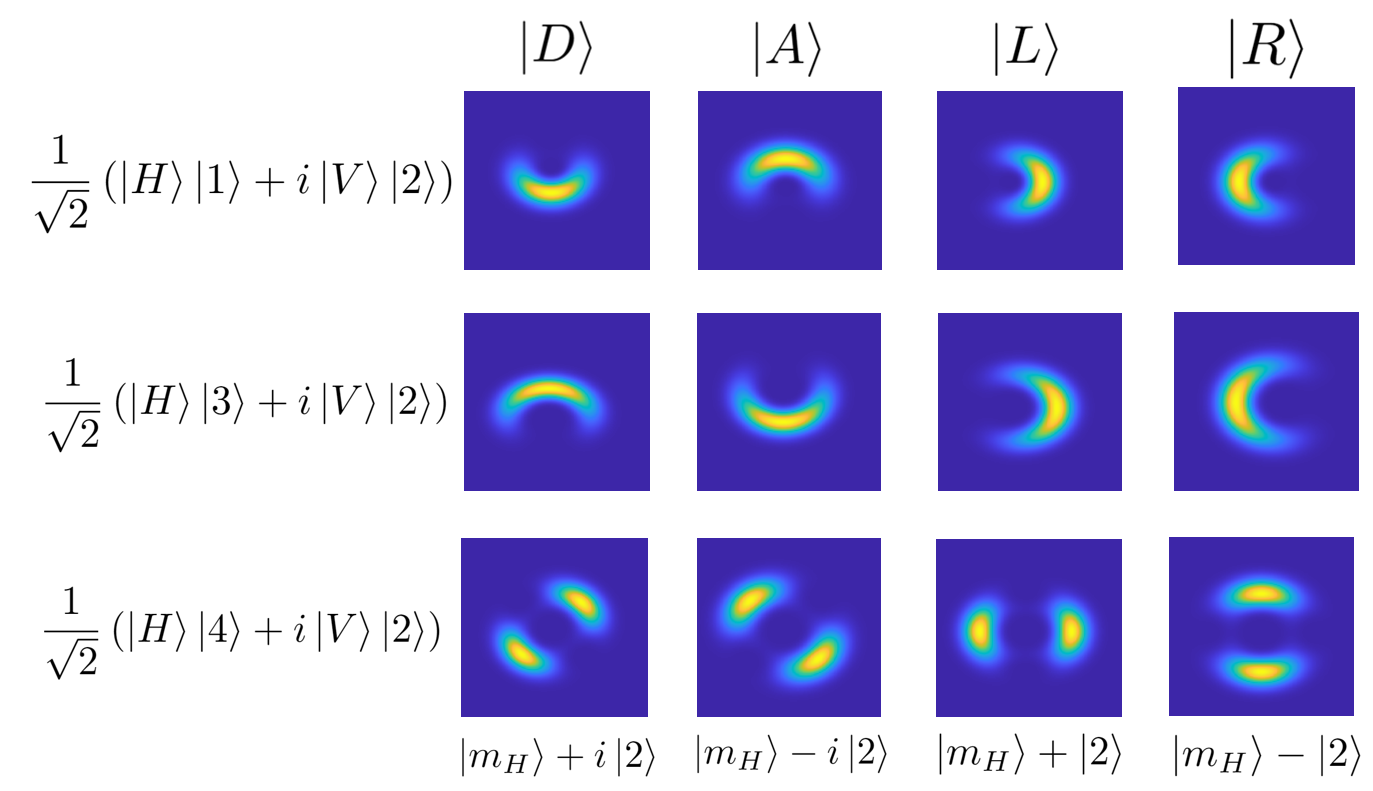}
    \caption{ Theoretically simulated intensity distributions of classical non-separable states for different polarization projections (linear (D/A) and circular (L/R)). SPP of order $m=2$ is used here and $m_H$ is the OAM value associated with horizontally polarized light.}
    \label{theory:cnss}
\end{figure}

 In this experiment, we used SPP of order $|m_{\text{SPP}}|=2$. We generated various non-separable states by changing the order of LG mode, $m_{\text{SLM}}$, with the help of SLM. For the detection of the state, a combination of QWP, HWP, and PBS is used. The combination of HWP and PBS acts as a polarizer. We measure the spatial distribution of light beam by projecting it to different polarizations such as linear polarization ($H$, $V$, $D$, and $A$) and circular polarization ($R$ and $L$). The intensity distribution for various non-separable states is given in Figures \ref{c7_11} and \ref{exp:cnss}. The theoretical simulated intensity distribution for various non-separable states are shown in Figures \ref{fig5} and \ref{theory:cnss}. Our experimental results shown in Figures \ref{c7_11} and \ref{exp:cnss} are in good agreement with the theoretical ones shown in Figures \ref{fig5} and \ref{theory:cnss} respectively. If we compare the experimental results with the theoretical simulation, it is very evident that SLM does not affect the spatial profile of vertical polarization. However, it only adds a relative phase delay of $\pi/2$ between $H$ and $V$ polarization due to the birefringent walkoff inside the SLM. \par
We also measured the Stokes parameters to calculate the degree of polarization. It is defined as,
\begin{align}
    S_0&=I_H+I_V,\hspace{30pt}S_1=I_D-I_A\nonumber \\
    S_2&=I_L-I_R ,\hspace{30pt}  S_3=I_H-I_V,
    \label{ch7_8}
\end{align}
where $I_x$ is the intensity of x-polarized light beam. The degree of polarization (DOP) can be written in terms of Stokes parameter,
\begin{equation}
    DOP=\sqrt{S_1^2+S_2^2+S_3^2},
\end{equation}
The $DOP$ ranges from 0, corresponding to the completely mixed polarized state (unpolarized light), to 1 for the completely polarized state. To characterize the non-separability of the state, linear entropy ($S_L$) can also be calculated. It can be represented in terms of $DOP$ \cite{perumangatt2015scattering},
\begin{equation}
    S_L= 1-DOP^2.
\end{equation}
The linear entropy measures the amount of mixedness present in the state. For a maximally entangled/non-separable system, the individual subsystem will always be in a mixed state. The maximum amount of mixedness present in the subsystem leads to the maximum non-separability of the system. Thus, one can measure the degree of non-separability by measuring the linear entropy $S_L$ of the subsystem. The linear entropy $S_L$ can range from 0, corresponding to the product state, and 1, corresponding to the maximally non-separable state.\par
 The values of $DOP$ and $S_L$ are given in Table \ref{tab7:0} for separable and non-separable states.
Without SPP and $m_{\text{SLM}}=0$ (Eqn. \ref{cc7_4}), the light beam is just a superposition of two orthogonal polarizations with Gaussian mode, which results in a completely polarized state (separable state). That is why, $DOP$ is maximum (0.94) without SPP and the linear entropy is $S_L=0.12$, which represents the product state of polarization and Gaussian mode.
\begin{table}[h!]
\centering
\begin{tabular}{ccc}
\hline
State &\hspace{20pt} $DOP$ & \hspace{20pt} Linear entropy $S_L$ \\
\hline 
Separable state  &\hspace{20pt} 0.94 &\hspace{10pt}0.12 \\
Non-separable state &\hspace{20pt} 0.05 &\hspace{10pt} 0.99  \\
\hline     
\end{tabular}
\caption{Experimentally recorded parameters, $DOP$ and $S_L$ for classical separable state and non-separable state.}
\label{tab7:0}
\end{table}
When we introduce the SPP and $m_{\text{SLM}}\neq0$, the $H$ and $V$ polarized components of light will correspond to different LG modes. $H$ polarized light is associated with LG mode of order $\ket{m_{\text{SPP}}+m_{\text{SLM}}}$ whereas $V$ polarized is associated with LG mode of order $\ket{m_{\text{SPP}}}$ (Eqn. \ref{cc7_4}). In this case, the state is completely unpolarized (or mixed polarized). Therefore the DOP is minimum (0.05) and the linear entropy is maximum (0.99) which shows the non-separability of the state in polarization and LG mode. We calculated the $DOP$ and linear entropy for each state shown in figure \ref{c7_11} and\ref{exp:cnss}. Since for each state we recorded more or less same $DOP$ and linear entropy, therefore for simplicity, we shown the value of $DOP$ and linear entropy of state $1/\sqrt{2}(\ket{H}\ket{-2}+e^{i\phi}\ket{V}\ket{2})$ in Table \ref{tab7:0}. We also recorded the spatial distribution of the beam by projecting it to the different polarization states (Figure \ref{c7_11} and \ref{exp:cnss} ). Due to the non-separability of the state, OAM state of the beam varies with projection to different polarization states. Since we used SPP of order $m=2$, that's why the OAM associated with $V$ polarized light is fixed. In order to achieve completely arbitrary NS state, one can either use SPP of higher order or replace the SPP with another SLM along with one HWP rotated at $45^{\circ}$ as explained in section \ref{sec2}.

\section{\label{sec4}Conclusion}
\noindent In conclusion, we found a new method to generate classical non-separable states using polarization and OAM degrees of freedom. The new setup is simple as compared to other interferometry methods, and it does not require any fine-tuning of alignment. We verified the presence of non-separablity by measuring the degree of polarization and linear entropy. The results are also in good agreement with the theory. \par
Since we could simultaneously encode polarization and OAM into a single photon, it will increase the information capacity per photon. Such a higher information capacity leads to higher secret key rates in quantum key distribution protocols. In this context, these hybrid states can be used as a powerful resource in many classical and quantum applications, such as in quantum communication, metrology, and quantum key distribution using quantum hybrid entangled states in particular.


\begin{thebibliography}{54}%
\makeatletter
\providecommand \@ifxundefined [1]{%
 \@ifx{#1\undefined}
}%
\providecommand \@ifnum [1]{%
 \ifnum #1\expandafter \@firstoftwo
 \else \expandafter \@secondoftwo
 \fi
}%
\providecommand \@ifx [1]{%
 \ifx #1\expandafter \@firstoftwo
 \else \expandafter \@secondoftwo
 \fi
}%
\providecommand \natexlab [1]{#1}%
\providecommand \enquote  [1]{``#1''}%
\providecommand \bibnamefont  [1]{#1}%
\providecommand \bibfnamefont [1]{#1}%
\providecommand \citenamefont [1]{#1}%
\providecommand \href@noop [0]{\@secondoftwo}%
\providecommand \href [0]{\begingroup \@sanitize@url \@href}%
\providecommand \@href[1]{\@@startlink{#1}\@@href}%
\providecommand \@@href[1]{\endgroup#1\@@endlink}%
\providecommand \@sanitize@url [0]{\catcode `\\12\catcode `\$12\catcode
  `\&12\catcode `\#12\catcode `\^12\catcode `\_12\catcode `\%12\relax}%
\providecommand \@@startlink[1]{}%
\providecommand \@@endlink[0]{}%
\providecommand \url  [0]{\begingroup\@sanitize@url \@url }%
\providecommand \@url [1]{\endgroup\@href {#1}{\urlprefix }}%
\providecommand \urlprefix  [0]{URL }%
\providecommand \Eprint [0]{\href }%
\providecommand \doibase [0]{https://doi.org/}%
\providecommand \selectlanguage [0]{\@gobble}%
\providecommand \bibinfo  [0]{\@secondoftwo}%
\providecommand \bibfield  [0]{\@secondoftwo}%
\providecommand \translation [1]{[#1]}%
\providecommand \BibitemOpen [0]{}%
\providecommand \bibitemStop [0]{}%
\providecommand \bibitemNoStop [0]{.\EOS\space}%
\providecommand \EOS [0]{\spacefactor3000\relax}%
\providecommand \BibitemShut  [1]{\csname bibitem#1\endcsname}%
\let\auto@bib@innerbib\@empty
\bibitem [{\citenamefont {Horodecki}\ \emph {et~al.}(2009)\citenamefont
  {Horodecki}, \citenamefont {Horodecki}, \citenamefont {Horodecki},\ and\
  \citenamefont {Horodecki}}]{horodecki2009quantum}%
  \BibitemOpen
  \bibfield  {author} {\bibinfo {author} {\bibfnamefont {R.}~\bibnamefont
  {Horodecki}}, \bibinfo {author} {\bibfnamefont {P.}~\bibnamefont
  {Horodecki}}, \bibinfo {author} {\bibfnamefont {M.}~\bibnamefont
  {Horodecki}},\ and\ \bibinfo {author} {\bibfnamefont {K.}~\bibnamefont
  {Horodecki}},\ }\bibfield  {title} {\bibinfo {title} {Quantum entanglement},\
  }\href@noop {} {\bibfield  {journal} {\bibinfo  {journal} {Reviews of modern
  physics}\ }\textbf {\bibinfo {volume} {81}},\ \bibinfo {pages} {865}
  (\bibinfo {year} {2009})}\BibitemShut {NoStop}%
\bibitem [{\citenamefont {Terhal}(2002)}]{terhal2002detecting}%
  \BibitemOpen
  \bibfield  {author} {\bibinfo {author} {\bibfnamefont {B.~M.}\ \bibnamefont
  {Terhal}},\ }\bibfield  {title} {\bibinfo {title} {Detecting quantum
  entanglement},\ }\href@noop {} {\bibfield  {journal} {\bibinfo  {journal}
  {Theoretical Computer Science}\ }\textbf {\bibinfo {volume} {287}},\ \bibinfo
  {pages} {313} (\bibinfo {year} {2002})}\BibitemShut {NoStop}%
\bibitem [{\citenamefont {Neves}\ \emph {et~al.}(2009)\citenamefont {Neves},
  \citenamefont {Lima}, \citenamefont {Delgado},\ and\ \citenamefont
  {Saavedra}}]{neves2009hybrid}%
  \BibitemOpen
  \bibfield  {author} {\bibinfo {author} {\bibfnamefont {L.}~\bibnamefont
  {Neves}}, \bibinfo {author} {\bibfnamefont {G.}~\bibnamefont {Lima}},
  \bibinfo {author} {\bibfnamefont {A.}~\bibnamefont {Delgado}},\ and\ \bibinfo
  {author} {\bibfnamefont {C.}~\bibnamefont {Saavedra}},\ }\bibfield  {title}
  {\bibinfo {title} {Hybrid photonic entanglement: Realization,
  characterization, and applications},\ }\href@noop {} {\bibfield  {journal}
  {\bibinfo  {journal} {Physical Review A}\ }\textbf {\bibinfo {volume} {80}},\
  \bibinfo {pages} {042322} (\bibinfo {year} {2009})}\BibitemShut {NoStop}%
\bibitem [{\citenamefont {Nagali}\ and\ \citenamefont
  {Sciarrino}(2010)}]{nagali2010generation}%
  \BibitemOpen
  \bibfield  {author} {\bibinfo {author} {\bibfnamefont {E.}~\bibnamefont
  {Nagali}}\ and\ \bibinfo {author} {\bibfnamefont {F.}~\bibnamefont
  {Sciarrino}},\ }\bibfield  {title} {\bibinfo {title} {Generation of hybrid
  polarization-orbital angular momentum entangled states},\ }\href@noop {}
  {\bibfield  {journal} {\bibinfo  {journal} {Optics express}\ }\textbf
  {\bibinfo {volume} {18}},\ \bibinfo {pages} {18243} (\bibinfo {year}
  {2010})}\BibitemShut {NoStop}%
\bibitem [{\citenamefont {Shen}\ and\ \citenamefont
  {Rosales-Guzm{\'a}n}(2022)}]{shen2022nonseparable}%
  \BibitemOpen
  \bibfield  {author} {\bibinfo {author} {\bibfnamefont {Y.}~\bibnamefont
  {Shen}}\ and\ \bibinfo {author} {\bibfnamefont {C.}~\bibnamefont
  {Rosales-Guzm{\'a}n}},\ }\bibfield  {title} {\bibinfo {title} {Nonseparable
  states of light: from quantum to classical},\ }\href@noop {} {\bibfield
  {journal} {\bibinfo  {journal} {Laser \& Photonics Reviews}\ }\textbf
  {\bibinfo {volume} {16}},\ \bibinfo {pages} {2100533} (\bibinfo {year}
  {2022})}\BibitemShut {NoStop}%
\bibitem [{\citenamefont {Kwiat}\ \emph {et~al.}(1995)\citenamefont {Kwiat},
  \citenamefont {Mattle}, \citenamefont {Weinfurter}, \citenamefont
  {Zeilinger}, \citenamefont {Sergienko},\ and\ \citenamefont
  {Shih}}]{kwiat1995new}%
  \BibitemOpen
  \bibfield  {author} {\bibinfo {author} {\bibfnamefont {P.~G.}\ \bibnamefont
  {Kwiat}}, \bibinfo {author} {\bibfnamefont {K.}~\bibnamefont {Mattle}},
  \bibinfo {author} {\bibfnamefont {H.}~\bibnamefont {Weinfurter}}, \bibinfo
  {author} {\bibfnamefont {A.}~\bibnamefont {Zeilinger}}, \bibinfo {author}
  {\bibfnamefont {A.~V.}\ \bibnamefont {Sergienko}},\ and\ \bibinfo {author}
  {\bibfnamefont {Y.}~\bibnamefont {Shih}},\ }\bibfield  {title} {\bibinfo
  {title} {New high-intensity source of polarization-entangled photon pairs},\
  }\href@noop {} {\bibfield  {journal} {\bibinfo  {journal} {Physical Review
  Letters}\ }\textbf {\bibinfo {volume} {75}},\ \bibinfo {pages} {4337}
  (\bibinfo {year} {1995})}\BibitemShut {NoStop}%
\bibitem [{\citenamefont {Dada}\ \emph {et~al.}(2011)\citenamefont {Dada},
  \citenamefont {Leach}, \citenamefont {Buller}, \citenamefont {Padgett},\ and\
  \citenamefont {Andersson}}]{dada2011experimental}%
  \BibitemOpen
  \bibfield  {author} {\bibinfo {author} {\bibfnamefont {A.~C.}\ \bibnamefont
  {Dada}}, \bibinfo {author} {\bibfnamefont {J.}~\bibnamefont {Leach}},
  \bibinfo {author} {\bibfnamefont {G.~S.}\ \bibnamefont {Buller}}, \bibinfo
  {author} {\bibfnamefont {M.~J.}\ \bibnamefont {Padgett}},\ and\ \bibinfo
  {author} {\bibfnamefont {E.}~\bibnamefont {Andersson}},\ }\bibfield  {title}
  {\bibinfo {title} {Experimental high-dimensional two-photon entanglement and
  violations of generalized bell inequalities},\ }\href@noop {} {\bibfield
  {journal} {\bibinfo  {journal} {Nature Physics}\ }\textbf {\bibinfo {volume}
  {7}},\ \bibinfo {pages} {677} (\bibinfo {year} {2011})}\BibitemShut {NoStop}%
\bibitem [{\citenamefont {Hamel}\ \emph {et~al.}(2014)\citenamefont {Hamel},
  \citenamefont {Shalm}, \citenamefont {H{\"u}bel}, \citenamefont {Miller},
  \citenamefont {Marsili}, \citenamefont {Verma}, \citenamefont {Mirin},
  \citenamefont {Nam}, \citenamefont {Resch},\ and\ \citenamefont
  {Jennewein}}]{hamel2014direct}%
  \BibitemOpen
  \bibfield  {author} {\bibinfo {author} {\bibfnamefont {D.~R.}\ \bibnamefont
  {Hamel}}, \bibinfo {author} {\bibfnamefont {L.~K.}\ \bibnamefont {Shalm}},
  \bibinfo {author} {\bibfnamefont {H.}~\bibnamefont {H{\"u}bel}}, \bibinfo
  {author} {\bibfnamefont {A.~J.}\ \bibnamefont {Miller}}, \bibinfo {author}
  {\bibfnamefont {F.}~\bibnamefont {Marsili}}, \bibinfo {author} {\bibfnamefont
  {V.~B.}\ \bibnamefont {Verma}}, \bibinfo {author} {\bibfnamefont {R.~P.}\
  \bibnamefont {Mirin}}, \bibinfo {author} {\bibfnamefont {S.~W.}\ \bibnamefont
  {Nam}}, \bibinfo {author} {\bibfnamefont {K.~J.}\ \bibnamefont {Resch}},\
  and\ \bibinfo {author} {\bibfnamefont {T.}~\bibnamefont {Jennewein}},\
  }\bibfield  {title} {\bibinfo {title} {Direct generation of three-photon
  polarization entanglement},\ }\href@noop {} {\bibfield  {journal} {\bibinfo
  {journal} {Nature Photonics}\ }\textbf {\bibinfo {volume} {8}},\ \bibinfo
  {pages} {801} (\bibinfo {year} {2014})}\BibitemShut {NoStop}%
\bibitem [{\citenamefont {Chithrabhanu}\ \emph {et~al.}(2016)\citenamefont
  {Chithrabhanu}, \citenamefont {Reddy}, \citenamefont {Lal}, \citenamefont
  {Anwar}, \citenamefont {Aadhi},\ and\ \citenamefont
  {Singh}}]{chithrabhanu2016pancharatnam}%
  \BibitemOpen
  \bibfield  {author} {\bibinfo {author} {\bibfnamefont {P.}~\bibnamefont
  {Chithrabhanu}}, \bibinfo {author} {\bibfnamefont {S.~G.}\ \bibnamefont
  {Reddy}}, \bibinfo {author} {\bibfnamefont {N.}~\bibnamefont {Lal}}, \bibinfo
  {author} {\bibfnamefont {A.}~\bibnamefont {Anwar}}, \bibinfo {author}
  {\bibfnamefont {A.}~\bibnamefont {Aadhi}},\ and\ \bibinfo {author}
  {\bibfnamefont {R.}~\bibnamefont {Singh}},\ }\bibfield  {title} {\bibinfo
  {title} {Pancharatnam phase in non-separable states of light},\ }\href@noop
  {} {\bibfield  {journal} {\bibinfo  {journal} {JOSA B}\ }\textbf {\bibinfo
  {volume} {33}},\ \bibinfo {pages} {2093} (\bibinfo {year}
  {2016})}\BibitemShut {NoStop}%
\bibitem [{\citenamefont {Spreeuw}(1998)}]{spreeuw1998classical}%
  \BibitemOpen
  \bibfield  {author} {\bibinfo {author} {\bibfnamefont {R.~J.}\ \bibnamefont
  {Spreeuw}},\ }\bibfield  {title} {\bibinfo {title} {A classical analogy of
  entanglement},\ }\href@noop {} {\bibfield  {journal} {\bibinfo  {journal}
  {Foundations of physics}\ }\textbf {\bibinfo {volume} {28}},\ \bibinfo
  {pages} {361} (\bibinfo {year} {1998})}\BibitemShut {NoStop}%
\bibitem [{\citenamefont {Ghose}\ and\ \citenamefont
  {Mukherjee}(2014)}]{ghose2014entanglement}%
  \BibitemOpen
  \bibfield  {author} {\bibinfo {author} {\bibfnamefont {P.}~\bibnamefont
  {Ghose}}\ and\ \bibinfo {author} {\bibfnamefont {A.}~\bibnamefont
  {Mukherjee}},\ }\bibfield  {title} {\bibinfo {title} {Entanglement in
  classical optics},\ }\href@noop {} {\bibfield  {journal} {\bibinfo  {journal}
  {Reviews in Theoretical Science}\ }\textbf {\bibinfo {volume} {2}},\ \bibinfo
  {pages} {274} (\bibinfo {year} {2014})}\BibitemShut {NoStop}%
\bibitem [{\citenamefont {Karimi}\ and\ \citenamefont
  {Boyd}(2015)}]{karimi2015classical}%
  \BibitemOpen
  \bibfield  {author} {\bibinfo {author} {\bibfnamefont {E.}~\bibnamefont
  {Karimi}}\ and\ \bibinfo {author} {\bibfnamefont {R.~W.}\ \bibnamefont
  {Boyd}},\ }\bibfield  {title} {\bibinfo {title} {Classical entanglement?},\
  }\href@noop {} {\bibfield  {journal} {\bibinfo  {journal} {Science}\ }\textbf
  {\bibinfo {volume} {350}},\ \bibinfo {pages} {1172} (\bibinfo {year}
  {2015})}\BibitemShut {NoStop}%
\bibitem [{\citenamefont {Paneru}\ \emph {et~al.}(2020)\citenamefont {Paneru},
  \citenamefont {Cohen}, \citenamefont {Fickler}, \citenamefont {Boyd},\ and\
  \citenamefont {Karimi}}]{paneru2020entanglement}%
  \BibitemOpen
  \bibfield  {author} {\bibinfo {author} {\bibfnamefont {D.}~\bibnamefont
  {Paneru}}, \bibinfo {author} {\bibfnamefont {E.}~\bibnamefont {Cohen}},
  \bibinfo {author} {\bibfnamefont {R.}~\bibnamefont {Fickler}}, \bibinfo
  {author} {\bibfnamefont {R.~W.}\ \bibnamefont {Boyd}},\ and\ \bibinfo
  {author} {\bibfnamefont {E.}~\bibnamefont {Karimi}},\ }\bibfield  {title}
  {\bibinfo {title} {Entanglement: quantum or classical?},\ }\href@noop {}
  {\bibfield  {journal} {\bibinfo  {journal} {Reports on Progress in Physics}\
  }\textbf {\bibinfo {volume} {83}},\ \bibinfo {pages} {064001} (\bibinfo
  {year} {2020})}\BibitemShut {NoStop}%
\bibitem [{\citenamefont {Luis}(2009)}]{luis2009coherence}%
  \BibitemOpen
  \bibfield  {author} {\bibinfo {author} {\bibfnamefont {A.}~\bibnamefont
  {Luis}},\ }\bibfield  {title} {\bibinfo {title} {Coherence, polarization, and
  entanglement for classical light fields},\ }\href@noop {} {\bibfield
  {journal} {\bibinfo  {journal} {Optics Communications}\ }\textbf {\bibinfo
  {volume} {282}},\ \bibinfo {pages} {3665} (\bibinfo {year}
  {2009})}\BibitemShut {NoStop}%
\bibitem [{\citenamefont {Borges}\ \emph {et~al.}(2010)\citenamefont {Borges},
  \citenamefont {Hor-Meyll}, \citenamefont {Huguenin},\ and\ \citenamefont
  {Khoury}}]{borges2010bell}%
  \BibitemOpen
  \bibfield  {author} {\bibinfo {author} {\bibfnamefont {C.}~\bibnamefont
  {Borges}}, \bibinfo {author} {\bibfnamefont {M.}~\bibnamefont {Hor-Meyll}},
  \bibinfo {author} {\bibfnamefont {J.}~\bibnamefont {Huguenin}},\ and\
  \bibinfo {author} {\bibfnamefont {A.}~\bibnamefont {Khoury}},\ }\bibfield
  {title} {\bibinfo {title} {Bell-like inequality for the spin-orbit
  separability of a laser beam},\ }\href@noop {} {\bibfield  {journal}
  {\bibinfo  {journal} {Physical Review A}\ }\textbf {\bibinfo {volume} {82}},\
  \bibinfo {pages} {033833} (\bibinfo {year} {2010})}\BibitemShut {NoStop}%
\bibitem [{\citenamefont {Pires}\ \emph {et~al.}(2010)\citenamefont {Pires},
  \citenamefont {Florijn},\ and\ \citenamefont
  {Van~Exter}}]{pires2010measurement}%
  \BibitemOpen
  \bibfield  {author} {\bibinfo {author} {\bibfnamefont {H.~D.~L.}\
  \bibnamefont {Pires}}, \bibinfo {author} {\bibfnamefont {H.}~\bibnamefont
  {Florijn}},\ and\ \bibinfo {author} {\bibfnamefont {M.}~\bibnamefont
  {Van~Exter}},\ }\bibfield  {title} {\bibinfo {title} {Measurement of the
  spiral spectrum of entangled two-photon states},\ }\href@noop {} {\bibfield
  {journal} {\bibinfo  {journal} {Physical review letters}\ }\textbf {\bibinfo
  {volume} {104}},\ \bibinfo {pages} {020505} (\bibinfo {year}
  {2010})}\BibitemShut {NoStop}%
\bibitem [{\citenamefont {Pereira}\ \emph {et~al.}(2014)\citenamefont
  {Pereira}, \citenamefont {Khoury},\ and\ \citenamefont
  {Dechoum}}]{pereira2014quantum}%
  \BibitemOpen
  \bibfield  {author} {\bibinfo {author} {\bibfnamefont {L.}~\bibnamefont
  {Pereira}}, \bibinfo {author} {\bibfnamefont {A.}~\bibnamefont {Khoury}},\
  and\ \bibinfo {author} {\bibfnamefont {K.}~\bibnamefont {Dechoum}},\
  }\bibfield  {title} {\bibinfo {title} {Quantum and classical separability of
  spin-orbit laser modes},\ }\href@noop {} {\bibfield  {journal} {\bibinfo
  {journal} {Physical Review A}\ }\textbf {\bibinfo {volume} {90}},\ \bibinfo
  {pages} {053842} (\bibinfo {year} {2014})}\BibitemShut {NoStop}%
\bibitem [{\citenamefont {Perumangatt}\ \emph {et~al.}(2015)\citenamefont
  {Perumangatt}, \citenamefont {Salla}, \citenamefont {Anwar}, \citenamefont
  {Aadhi}, \citenamefont {Prabhakar},\ and\ \citenamefont
  {Singh}}]{perumangatt2015scattering}%
  \BibitemOpen
  \bibfield  {author} {\bibinfo {author} {\bibfnamefont {C.}~\bibnamefont
  {Perumangatt}}, \bibinfo {author} {\bibfnamefont {G.~R.}\ \bibnamefont
  {Salla}}, \bibinfo {author} {\bibfnamefont {A.}~\bibnamefont {Anwar}},
  \bibinfo {author} {\bibfnamefont {A.}~\bibnamefont {Aadhi}}, \bibinfo
  {author} {\bibfnamefont {S.}~\bibnamefont {Prabhakar}},\ and\ \bibinfo
  {author} {\bibfnamefont {R.}~\bibnamefont {Singh}},\ }\bibfield  {title}
  {\bibinfo {title} {Scattering of non-separable states of light},\ }\href@noop
  {} {\bibfield  {journal} {\bibinfo  {journal} {Optics Communications}\
  }\textbf {\bibinfo {volume} {355}},\ \bibinfo {pages} {301} (\bibinfo {year}
  {2015})}\BibitemShut {NoStop}%
\bibitem [{\citenamefont {Salla}\ \emph {et~al.}(2015)\citenamefont {Salla},
  \citenamefont {Perumangattu}, \citenamefont {Prabhakar}, \citenamefont
  {Anwar},\ and\ \citenamefont {Singh}}]{salla2015recovering}%
  \BibitemOpen
  \bibfield  {author} {\bibinfo {author} {\bibfnamefont {G.~R.}\ \bibnamefont
  {Salla}}, \bibinfo {author} {\bibfnamefont {C.}~\bibnamefont {Perumangattu}},
  \bibinfo {author} {\bibfnamefont {S.}~\bibnamefont {Prabhakar}}, \bibinfo
  {author} {\bibfnamefont {A.}~\bibnamefont {Anwar}},\ and\ \bibinfo {author}
  {\bibfnamefont {R.~P.}\ \bibnamefont {Singh}},\ }\bibfield  {title} {\bibinfo
  {title} {Recovering the vorticity of a light beam after scattering},\
  }\href@noop {} {\bibfield  {journal} {\bibinfo  {journal} {Applied Physics
  Letters}\ }\textbf {\bibinfo {volume} {107}},\ \bibinfo {pages} {021104}
  (\bibinfo {year} {2015})}\BibitemShut {NoStop}%
\bibitem [{\citenamefont {Galvez}\ \emph {et~al.}(2012)\citenamefont {Galvez},
  \citenamefont {Khadka}, \citenamefont {Schubert},\ and\ \citenamefont
  {Nomoto}}]{galvez2012poincare}%
  \BibitemOpen
  \bibfield  {author} {\bibinfo {author} {\bibfnamefont {E.~J.}\ \bibnamefont
  {Galvez}}, \bibinfo {author} {\bibfnamefont {S.}~\bibnamefont {Khadka}},
  \bibinfo {author} {\bibfnamefont {W.~H.}\ \bibnamefont {Schubert}},\ and\
  \bibinfo {author} {\bibfnamefont {S.}~\bibnamefont {Nomoto}},\ }\bibfield
  {title} {\bibinfo {title} {Poincar{\'e}-beam patterns produced by
  nonseparable superpositions of laguerre--gauss and polarization modes of
  light},\ }\href@noop {} {\bibfield  {journal} {\bibinfo  {journal} {Applied
  optics}\ }\textbf {\bibinfo {volume} {51}},\ \bibinfo {pages} {2925}
  (\bibinfo {year} {2012})}\BibitemShut {NoStop}%
\bibitem [{\citenamefont {Beckley}\ \emph {et~al.}(2010)\citenamefont
  {Beckley}, \citenamefont {Brown},\ and\ \citenamefont
  {Alonso}}]{beckley2010full}%
  \BibitemOpen
  \bibfield  {author} {\bibinfo {author} {\bibfnamefont {A.~M.}\ \bibnamefont
  {Beckley}}, \bibinfo {author} {\bibfnamefont {T.~G.}\ \bibnamefont {Brown}},\
  and\ \bibinfo {author} {\bibfnamefont {M.~A.}\ \bibnamefont {Alonso}},\
  }\bibfield  {title} {\bibinfo {title} {Full poincar{\'e} beams},\ }\href@noop
  {} {\bibfield  {journal} {\bibinfo  {journal} {Optics express}\ }\textbf
  {\bibinfo {volume} {18}},\ \bibinfo {pages} {10777} (\bibinfo {year}
  {2010})}\BibitemShut {NoStop}%
\bibitem [{\citenamefont {Aiello}\ \emph {et~al.}(2015)\citenamefont {Aiello},
  \citenamefont {T{\"o}ppel}, \citenamefont {Marquardt}, \citenamefont
  {Giacobino},\ and\ \citenamefont {Leuchs}}]{aiello2015quantum}%
  \BibitemOpen
  \bibfield  {author} {\bibinfo {author} {\bibfnamefont {A.}~\bibnamefont
  {Aiello}}, \bibinfo {author} {\bibfnamefont {F.}~\bibnamefont {T{\"o}ppel}},
  \bibinfo {author} {\bibfnamefont {C.}~\bibnamefont {Marquardt}}, \bibinfo
  {author} {\bibfnamefont {E.}~\bibnamefont {Giacobino}},\ and\ \bibinfo
  {author} {\bibfnamefont {G.}~\bibnamefont {Leuchs}},\ }\bibfield  {title}
  {\bibinfo {title} {Quantum- like nonseparable structures in optical beams},\
  }\href@noop {} {\bibfield  {journal} {\bibinfo  {journal} {New Journal of
  Physics}\ }\textbf {\bibinfo {volume} {17}},\ \bibinfo {pages} {043024}
  (\bibinfo {year} {2015})}\BibitemShut {NoStop}%
\bibitem [{\citenamefont {Ndagano}\ \emph {et~al.}(2017)\citenamefont
  {Ndagano}, \citenamefont {Perez-Garcia}, \citenamefont {Roux}, \citenamefont
  {McLaren}, \citenamefont {Rosales-Guzman}, \citenamefont {Zhang},
  \citenamefont {Mouane}, \citenamefont {Hernandez-Aranda}, \citenamefont
  {Konrad},\ and\ \citenamefont {Forbes}}]{ndagano2017characterizing}%
  \BibitemOpen
  \bibfield  {author} {\bibinfo {author} {\bibfnamefont {B.}~\bibnamefont
  {Ndagano}}, \bibinfo {author} {\bibfnamefont {B.}~\bibnamefont
  {Perez-Garcia}}, \bibinfo {author} {\bibfnamefont {F.~S.}\ \bibnamefont
  {Roux}}, \bibinfo {author} {\bibfnamefont {M.}~\bibnamefont {McLaren}},
  \bibinfo {author} {\bibfnamefont {C.}~\bibnamefont {Rosales-Guzman}},
  \bibinfo {author} {\bibfnamefont {Y.}~\bibnamefont {Zhang}}, \bibinfo
  {author} {\bibfnamefont {O.}~\bibnamefont {Mouane}}, \bibinfo {author}
  {\bibfnamefont {R.~I.}\ \bibnamefont {Hernandez-Aranda}}, \bibinfo {author}
  {\bibfnamefont {T.}~\bibnamefont {Konrad}},\ and\ \bibinfo {author}
  {\bibfnamefont {A.}~\bibnamefont {Forbes}},\ }\bibfield  {title} {\bibinfo
  {title} {Characterizing quantum channels with non-separable states of
  classical light},\ }\href@noop {} {\bibfield  {journal} {\bibinfo  {journal}
  {Nature Physics}\ }\textbf {\bibinfo {volume} {13}},\ \bibinfo {pages} {397}
  (\bibinfo {year} {2017})}\BibitemShut {NoStop}%
\bibitem [{\citenamefont {Ndagano}\ \emph {et~al.}(2018)\citenamefont
  {Ndagano}, \citenamefont {Nape}, \citenamefont {Cox}, \citenamefont
  {Rosales-Guzman},\ and\ \citenamefont {Forbes}}]{ndagano2018creation}%
  \BibitemOpen
  \bibfield  {author} {\bibinfo {author} {\bibfnamefont {B.}~\bibnamefont
  {Ndagano}}, \bibinfo {author} {\bibfnamefont {I.}~\bibnamefont {Nape}},
  \bibinfo {author} {\bibfnamefont {M.~A.}\ \bibnamefont {Cox}}, \bibinfo
  {author} {\bibfnamefont {C.}~\bibnamefont {Rosales-Guzman}},\ and\ \bibinfo
  {author} {\bibfnamefont {A.}~\bibnamefont {Forbes}},\ }\bibfield  {title}
  {\bibinfo {title} {Creation and detection of vector vortex modes for
  classical and quantum communication},\ }\href@noop {} {\bibfield  {journal}
  {\bibinfo  {journal} {Journal of Lightwave Technology}\ }\textbf {\bibinfo
  {volume} {36}},\ \bibinfo {pages} {292} (\bibinfo {year} {2018})}\BibitemShut
  {NoStop}%
\bibitem [{\citenamefont {Milione}\ \emph {et~al.}(2015)\citenamefont
  {Milione}, \citenamefont {Nguyen}, \citenamefont {Leach}, \citenamefont
  {Nolan},\ and\ \citenamefont {Alfano}}]{milione2015using}%
  \BibitemOpen
  \bibfield  {author} {\bibinfo {author} {\bibfnamefont {G.}~\bibnamefont
  {Milione}}, \bibinfo {author} {\bibfnamefont {T.~A.}\ \bibnamefont {Nguyen}},
  \bibinfo {author} {\bibfnamefont {J.}~\bibnamefont {Leach}}, \bibinfo
  {author} {\bibfnamefont {D.~A.}\ \bibnamefont {Nolan}},\ and\ \bibinfo
  {author} {\bibfnamefont {R.~R.}\ \bibnamefont {Alfano}},\ }\bibfield  {title}
  {\bibinfo {title} {Using the nonseparability of vector beams to encode
  information for optical communication},\ }\href@noop {} {\bibfield  {journal}
  {\bibinfo  {journal} {Optics letters}\ }\textbf {\bibinfo {volume} {40}},\
  \bibinfo {pages} {4887} (\bibinfo {year} {2015})}\BibitemShut {NoStop}%
\bibitem [{\citenamefont {Krenn}\ \emph {et~al.}(2015)\citenamefont {Krenn},
  \citenamefont {Handsteiner}, \citenamefont {Fink}, \citenamefont {Fickler},\
  and\ \citenamefont {Zeilinger}}]{krenn2015twisted}%
  \BibitemOpen
  \bibfield  {author} {\bibinfo {author} {\bibfnamefont {M.}~\bibnamefont
  {Krenn}}, \bibinfo {author} {\bibfnamefont {J.}~\bibnamefont {Handsteiner}},
  \bibinfo {author} {\bibfnamefont {M.}~\bibnamefont {Fink}}, \bibinfo {author}
  {\bibfnamefont {R.}~\bibnamefont {Fickler}},\ and\ \bibinfo {author}
  {\bibfnamefont {A.}~\bibnamefont {Zeilinger}},\ }\bibfield  {title} {\bibinfo
  {title} {Twisted photon entanglement through turbulent air across vienna},\
  }\href@noop {} {\bibfield  {journal} {\bibinfo  {journal} {Proceedings of the
  National Academy of Sciences}\ }\textbf {\bibinfo {volume} {112}},\ \bibinfo
  {pages} {14197} (\bibinfo {year} {2015})}\BibitemShut {NoStop}%
\bibitem [{\citenamefont {Otte}\ \emph {et~al.}(2018)\citenamefont {Otte},
  \citenamefont {Rosales-Guzm{\'a}n}, \citenamefont {Ndagano}, \citenamefont
  {Denz},\ and\ \citenamefont {Forbes}}]{otte2018entanglement}%
  \BibitemOpen
  \bibfield  {author} {\bibinfo {author} {\bibfnamefont {E.}~\bibnamefont
  {Otte}}, \bibinfo {author} {\bibfnamefont {C.}~\bibnamefont
  {Rosales-Guzm{\'a}n}}, \bibinfo {author} {\bibfnamefont {B.}~\bibnamefont
  {Ndagano}}, \bibinfo {author} {\bibfnamefont {C.}~\bibnamefont {Denz}},\ and\
  \bibinfo {author} {\bibfnamefont {A.}~\bibnamefont {Forbes}},\ }\bibfield
  {title} {\bibinfo {title} {Entanglement beating in free space through
  spin--orbit coupling},\ }\href@noop {} {\bibfield  {journal} {\bibinfo
  {journal} {Light: Science \& Applications}\ }\textbf {\bibinfo {volume}
  {7}},\ \bibinfo {pages} {18009} (\bibinfo {year} {2018})}\BibitemShut
  {NoStop}%
\bibitem [{\citenamefont {T{\"o}ppel}\ \emph {et~al.}(2014)\citenamefont
  {T{\"o}ppel}, \citenamefont {Aiello}, \citenamefont {Marquardt},
  \citenamefont {Giacobino},\ and\ \citenamefont
  {Leuchs}}]{toppel2014classical}%
  \BibitemOpen
  \bibfield  {author} {\bibinfo {author} {\bibfnamefont {F.}~\bibnamefont
  {T{\"o}ppel}}, \bibinfo {author} {\bibfnamefont {A.}~\bibnamefont {Aiello}},
  \bibinfo {author} {\bibfnamefont {C.}~\bibnamefont {Marquardt}}, \bibinfo
  {author} {\bibfnamefont {E.}~\bibnamefont {Giacobino}},\ and\ \bibinfo
  {author} {\bibfnamefont {G.}~\bibnamefont {Leuchs}},\ }\bibfield  {title}
  {\bibinfo {title} {Classical entanglement in polarization metrology},\
  }\href@noop {} {\bibfield  {journal} {\bibinfo  {journal} {New Journal of
  Physics}\ }\textbf {\bibinfo {volume} {16}},\ \bibinfo {pages} {073019}
  (\bibinfo {year} {2014})}\BibitemShut {NoStop}%
\bibitem [{\citenamefont {Maga{\~n}a-Loaiza}\ and\ \citenamefont
  {Boyd}(2019)}]{magana2019quantum}%
  \BibitemOpen
  \bibfield  {author} {\bibinfo {author} {\bibfnamefont {O.~S.}\ \bibnamefont
  {Maga{\~n}a-Loaiza}}\ and\ \bibinfo {author} {\bibfnamefont {R.~W.}\
  \bibnamefont {Boyd}},\ }\bibfield  {title} {\bibinfo {title} {Quantum imaging
  and information},\ }\href@noop {} {\bibfield  {journal} {\bibinfo  {journal}
  {Reports on Progress in Physics}\ }\textbf {\bibinfo {volume} {82}},\
  \bibinfo {pages} {124401} (\bibinfo {year} {2019})}\BibitemShut {NoStop}%
\bibitem [{\citenamefont {Daboul}\ \emph {et~al.}(2003)\citenamefont {Daboul},
  \citenamefont {Wang},\ and\ \citenamefont {Sanders}}]{daboul2003quantum}%
  \BibitemOpen
  \bibfield  {author} {\bibinfo {author} {\bibfnamefont {J.}~\bibnamefont
  {Daboul}}, \bibinfo {author} {\bibfnamefont {X.}~\bibnamefont {Wang}},\ and\
  \bibinfo {author} {\bibfnamefont {B.~C.}\ \bibnamefont {Sanders}},\
  }\bibfield  {title} {\bibinfo {title} {Quantum gates on hybrid qudits},\
  }\href@noop {} {\bibfield  {journal} {\bibinfo  {journal} {Journal of Physics
  A: Mathematical and General}\ }\textbf {\bibinfo {volume} {36}},\ \bibinfo
  {pages} {2525} (\bibinfo {year} {2003})}\BibitemShut {NoStop}%
\bibitem [{\citenamefont {Allen}\ \emph {et~al.}(1992)\citenamefont {Allen},
  \citenamefont {Beijersbergen}, \citenamefont {Spreeuw},\ and\ \citenamefont
  {Woerdman}}]{allen1992orbital}%
  \BibitemOpen
  \bibfield  {author} {\bibinfo {author} {\bibfnamefont {L.}~\bibnamefont
  {Allen}}, \bibinfo {author} {\bibfnamefont {M.~W.}\ \bibnamefont
  {Beijersbergen}}, \bibinfo {author} {\bibfnamefont {R.}~\bibnamefont
  {Spreeuw}},\ and\ \bibinfo {author} {\bibfnamefont {J.}~\bibnamefont
  {Woerdman}},\ }\bibfield  {title} {\bibinfo {title} {Orbital angular momentum
  of light and the transformation of laguerre-gaussian laser modes},\
  }\href@noop {} {\bibfield  {journal} {\bibinfo  {journal} {Physical review
  A}\ }\textbf {\bibinfo {volume} {45}},\ \bibinfo {pages} {8185} (\bibinfo
  {year} {1992})}\BibitemShut {NoStop}%
\bibitem [{\citenamefont {Bai}\ \emph {et~al.}(2022)\citenamefont {Bai},
  \citenamefont {Lv}, \citenamefont {Fu},\ and\ \citenamefont
  {Yang}}]{bai2022vortex}%
  \BibitemOpen
  \bibfield  {author} {\bibinfo {author} {\bibfnamefont {Y.}~\bibnamefont
  {Bai}}, \bibinfo {author} {\bibfnamefont {H.}~\bibnamefont {Lv}}, \bibinfo
  {author} {\bibfnamefont {X.}~\bibnamefont {Fu}},\ and\ \bibinfo {author}
  {\bibfnamefont {Y.}~\bibnamefont {Yang}},\ }\bibfield  {title} {\bibinfo
  {title} {Vortex beam: generation and detection of orbital angular momentum},\
  }\href@noop {} {\bibfield  {journal} {\bibinfo  {journal} {Chinese Optics
  Letters}\ }\textbf {\bibinfo {volume} {20}},\ \bibinfo {pages} {012601}
  (\bibinfo {year} {2022})}\BibitemShut {NoStop}%
\bibitem [{\citenamefont {Willner}\ \emph {et~al.}(2015)\citenamefont
  {Willner}, \citenamefont {Huang}, \citenamefont {Yan}, \citenamefont {Ren},
  \citenamefont {Ahmed}, \citenamefont {Xie}, \citenamefont {Bao},
  \citenamefont {Li}, \citenamefont {Cao}, \citenamefont {Zhao} \emph
  {et~al.}}]{willner2015optical}%
  \BibitemOpen
  \bibfield  {author} {\bibinfo {author} {\bibfnamefont {A.~E.}\ \bibnamefont
  {Willner}}, \bibinfo {author} {\bibfnamefont {H.}~\bibnamefont {Huang}},
  \bibinfo {author} {\bibfnamefont {Y.}~\bibnamefont {Yan}}, \bibinfo {author}
  {\bibfnamefont {Y.}~\bibnamefont {Ren}}, \bibinfo {author} {\bibfnamefont
  {N.}~\bibnamefont {Ahmed}}, \bibinfo {author} {\bibfnamefont
  {G.}~\bibnamefont {Xie}}, \bibinfo {author} {\bibfnamefont {C.}~\bibnamefont
  {Bao}}, \bibinfo {author} {\bibfnamefont {L.}~\bibnamefont {Li}}, \bibinfo
  {author} {\bibfnamefont {Y.}~\bibnamefont {Cao}}, \bibinfo {author}
  {\bibfnamefont {Z.}~\bibnamefont {Zhao}}, \emph {et~al.},\ }\bibfield
  {title} {\bibinfo {title} {Optical communications using orbital angular
  momentum beams},\ }\href@noop {} {\bibfield  {journal} {\bibinfo  {journal}
  {Advances in optics and photonics}\ }\textbf {\bibinfo {volume} {7}},\
  \bibinfo {pages} {66} (\bibinfo {year} {2015})}\BibitemShut {NoStop}%
\bibitem [{\citenamefont {Forbes}\ and\ \citenamefont
  {Nape}(2019)}]{forbes2019quantum}%
  \BibitemOpen
  \bibfield  {author} {\bibinfo {author} {\bibfnamefont {A.}~\bibnamefont
  {Forbes}}\ and\ \bibinfo {author} {\bibfnamefont {I.}~\bibnamefont {Nape}},\
  }\bibfield  {title} {\bibinfo {title} {Quantum mechanics with patterns of
  light: progress in high dimensional and multidimensional entanglement with
  structured light},\ }\href@noop {} {\bibfield  {journal} {\bibinfo  {journal}
  {AVS Quantum Science}\ }\textbf {\bibinfo {volume} {1}} (\bibinfo {year}
  {2019})}\BibitemShut {NoStop}%
\bibitem [{\citenamefont {Rozenberg}\ \emph {et~al.}(2023)\citenamefont
  {Rozenberg}, \citenamefont {Karnieli}, \citenamefont {Yesharim},
  \citenamefont {Foley-Comer}, \citenamefont {Trajtenberg-Mills}, \citenamefont
  {Mishra}, \citenamefont {Prabhakar}, \citenamefont {Pratap}, \citenamefont
  {Freedman}, \citenamefont {Bronstein} \emph
  {et~al.}}]{rozenberg2023designing}%
  \BibitemOpen
  \bibfield  {author} {\bibinfo {author} {\bibfnamefont {E.}~\bibnamefont
  {Rozenberg}}, \bibinfo {author} {\bibfnamefont {A.}~\bibnamefont {Karnieli}},
  \bibinfo {author} {\bibfnamefont {O.}~\bibnamefont {Yesharim}}, \bibinfo
  {author} {\bibfnamefont {J.}~\bibnamefont {Foley-Comer}}, \bibinfo {author}
  {\bibfnamefont {S.}~\bibnamefont {Trajtenberg-Mills}}, \bibinfo {author}
  {\bibfnamefont {S.}~\bibnamefont {Mishra}}, \bibinfo {author} {\bibfnamefont
  {S.}~\bibnamefont {Prabhakar}}, \bibinfo {author} {\bibfnamefont
  {R.}~\bibnamefont {Pratap}}, \bibinfo {author} {\bibfnamefont
  {D.}~\bibnamefont {Freedman}}, \bibinfo {author} {\bibfnamefont {A.~M.}\
  \bibnamefont {Bronstein}}, \emph {et~al.},\ }\bibfield  {title} {\bibinfo
  {title} {Designing nonlinear photonic crystals for high-dimensional quantum
  state engineering},\ }\href@noop {} {\bibfield  {journal} {\bibinfo
  {journal} {arXiv preprint arXiv:2304.06810}\ } (\bibinfo {year}
  {2023})}\BibitemShut {NoStop}%
\bibitem [{\citenamefont {Bacco}\ \emph {et~al.}(2020)\citenamefont {Bacco},
  \citenamefont {Cozzolino}, \citenamefont {Da~Lio}, \citenamefont {Ding},
  \citenamefont {Rottwitt},\ and\ \citenamefont
  {Oxenl{\o}we}}]{bacco2020quantum}%
  \BibitemOpen
  \bibfield  {author} {\bibinfo {author} {\bibfnamefont {D.}~\bibnamefont
  {Bacco}}, \bibinfo {author} {\bibfnamefont {D.}~\bibnamefont {Cozzolino}},
  \bibinfo {author} {\bibfnamefont {B.}~\bibnamefont {Da~Lio}}, \bibinfo
  {author} {\bibfnamefont {Y.}~\bibnamefont {Ding}}, \bibinfo {author}
  {\bibfnamefont {K.}~\bibnamefont {Rottwitt}},\ and\ \bibinfo {author}
  {\bibfnamefont {L.~K.}\ \bibnamefont {Oxenl{\o}we}},\ }\bibfield  {title}
  {\bibinfo {title} {Quantum communication with orbital angular momentum},\
  }in\ \href@noop {} {\emph {\bibinfo {booktitle} {2020 22nd International
  Conference on Transparent Optical Networks (ICTON)}}}\ (\bibinfo
  {organization} {IEEE},\ \bibinfo {year} {2020})\ pp.\ \bibinfo {pages}
  {1--4}\BibitemShut {NoStop}%
\bibitem [{\citenamefont {Cozzolino}\ \emph
  {et~al.}(2019{\natexlab{a}})\citenamefont {Cozzolino}, \citenamefont {Bacco},
  \citenamefont {Da~Lio}, \citenamefont {Ingerslev}, \citenamefont {Ding},
  \citenamefont {Dalgaard}, \citenamefont {Kristensen}, \citenamefont {Galili},
  \citenamefont {Rottwitt}, \citenamefont {Ramachandran} \emph
  {et~al.}}]{cozzolino2019orbital}%
  \BibitemOpen
  \bibfield  {author} {\bibinfo {author} {\bibfnamefont {D.}~\bibnamefont
  {Cozzolino}}, \bibinfo {author} {\bibfnamefont {D.}~\bibnamefont {Bacco}},
  \bibinfo {author} {\bibfnamefont {B.}~\bibnamefont {Da~Lio}}, \bibinfo
  {author} {\bibfnamefont {K.}~\bibnamefont {Ingerslev}}, \bibinfo {author}
  {\bibfnamefont {Y.}~\bibnamefont {Ding}}, \bibinfo {author} {\bibfnamefont
  {K.}~\bibnamefont {Dalgaard}}, \bibinfo {author} {\bibfnamefont
  {P.}~\bibnamefont {Kristensen}}, \bibinfo {author} {\bibfnamefont
  {M.}~\bibnamefont {Galili}}, \bibinfo {author} {\bibfnamefont
  {K.}~\bibnamefont {Rottwitt}}, \bibinfo {author} {\bibfnamefont
  {S.}~\bibnamefont {Ramachandran}}, \emph {et~al.},\ }\bibfield  {title}
  {\bibinfo {title} {Orbital angular momentum states enabling fiber-based
  high-dimensional quantum communication},\ }\href@noop {} {\bibfield
  {journal} {\bibinfo  {journal} {Physical Review Applied}\ }\textbf {\bibinfo
  {volume} {11}},\ \bibinfo {pages} {064058} (\bibinfo {year}
  {2019}{\natexlab{a}})}\BibitemShut {NoStop}%
\bibitem [{\citenamefont {Jabir}\ \emph {et~al.}(2017)\citenamefont {Jabir},
  \citenamefont {Apurv~Chaitanya}, \citenamefont {Mathew},\ and\ \citenamefont
  {Samanta}}]{jabir2017direct}%
  \BibitemOpen
  \bibfield  {author} {\bibinfo {author} {\bibfnamefont {M.}~\bibnamefont
  {Jabir}}, \bibinfo {author} {\bibfnamefont {N.}~\bibnamefont
  {Apurv~Chaitanya}}, \bibinfo {author} {\bibfnamefont {M.}~\bibnamefont
  {Mathew}},\ and\ \bibinfo {author} {\bibfnamefont {G.}~\bibnamefont
  {Samanta}},\ }\bibfield  {title} {\bibinfo {title} {Direct transfer of
  classical non-separable states into hybrid entangled two photon states},\
  }\href@noop {} {\bibfield  {journal} {\bibinfo  {journal} {Scientific
  Reports}\ }\textbf {\bibinfo {volume} {7}},\ \bibinfo {pages} {1} (\bibinfo
  {year} {2017})}\BibitemShut {NoStop}%
\bibitem [{\citenamefont {Cozzolino}\ \emph
  {et~al.}(2019{\natexlab{b}})\citenamefont {Cozzolino}, \citenamefont
  {Polino}, \citenamefont {Valeri}, \citenamefont {Carvacho}, \citenamefont
  {Bacco}, \citenamefont {Spagnolo}, \citenamefont {Oxenl{\o}we},\ and\
  \citenamefont {Sciarrino}}]{cozzolino2019air}%
  \BibitemOpen
  \bibfield  {author} {\bibinfo {author} {\bibfnamefont {D.}~\bibnamefont
  {Cozzolino}}, \bibinfo {author} {\bibfnamefont {E.}~\bibnamefont {Polino}},
  \bibinfo {author} {\bibfnamefont {M.}~\bibnamefont {Valeri}}, \bibinfo
  {author} {\bibfnamefont {G.}~\bibnamefont {Carvacho}}, \bibinfo {author}
  {\bibfnamefont {D.}~\bibnamefont {Bacco}}, \bibinfo {author} {\bibfnamefont
  {N.}~\bibnamefont {Spagnolo}}, \bibinfo {author} {\bibfnamefont {L.~K.}\
  \bibnamefont {Oxenl{\o}we}},\ and\ \bibinfo {author} {\bibfnamefont
  {F.}~\bibnamefont {Sciarrino}},\ }\bibfield  {title} {\bibinfo {title}
  {Air-core fiber distribution of hybrid vector vortex-polarization entangled
  states},\ }\href@noop {} {\bibfield  {journal} {\bibinfo  {journal} {Advanced
  Photonics}\ }\textbf {\bibinfo {volume} {1}},\ \bibinfo {pages} {046005}
  (\bibinfo {year} {2019}{\natexlab{b}})}\BibitemShut {NoStop}%
\bibitem [{\citenamefont {Slussarenko}\ \emph {et~al.}(2010)\citenamefont
  {Slussarenko}, \citenamefont {D’Ambrosio}, \citenamefont {Piccirillo},
  \citenamefont {Marrucci},\ and\ \citenamefont
  {Santamato}}]{slussarenko2010polarizing}%
  \BibitemOpen
  \bibfield  {author} {\bibinfo {author} {\bibfnamefont {S.}~\bibnamefont
  {Slussarenko}}, \bibinfo {author} {\bibfnamefont {V.}~\bibnamefont
  {D’Ambrosio}}, \bibinfo {author} {\bibfnamefont {B.}~\bibnamefont
  {Piccirillo}}, \bibinfo {author} {\bibfnamefont {L.}~\bibnamefont
  {Marrucci}},\ and\ \bibinfo {author} {\bibfnamefont {E.}~\bibnamefont
  {Santamato}},\ }\bibfield  {title} {\bibinfo {title} {The polarizing sagnac
  interferometer: a tool for light orbital angular momentum sorting and
  spin-orbit photon processing},\ }\href@noop {} {\bibfield  {journal}
  {\bibinfo  {journal} {Optics Express}\ }\textbf {\bibinfo {volume} {18}},\
  \bibinfo {pages} {27205} (\bibinfo {year} {2010})}\BibitemShut {NoStop}%
\bibitem [{\citenamefont {Maurer}\ \emph {et~al.}(2007)\citenamefont {Maurer},
  \citenamefont {Jesacher}, \citenamefont {F{\"u}rhapter}, \citenamefont
  {Bernet},\ and\ \citenamefont {Ritsch-Marte}}]{maurer2007tailoring}%
  \BibitemOpen
  \bibfield  {author} {\bibinfo {author} {\bibfnamefont {C.}~\bibnamefont
  {Maurer}}, \bibinfo {author} {\bibfnamefont {A.}~\bibnamefont {Jesacher}},
  \bibinfo {author} {\bibfnamefont {S.}~\bibnamefont {F{\"u}rhapter}}, \bibinfo
  {author} {\bibfnamefont {S.}~\bibnamefont {Bernet}},\ and\ \bibinfo {author}
  {\bibfnamefont {M.}~\bibnamefont {Ritsch-Marte}},\ }\bibfield  {title}
  {\bibinfo {title} {Tailoring of arbitrary optical vector beams},\ }\href@noop
  {} {\bibfield  {journal} {\bibinfo  {journal} {New Journal of Physics}\
  }\textbf {\bibinfo {volume} {9}},\ \bibinfo {pages} {78} (\bibinfo {year}
  {2007})}\BibitemShut {NoStop}%
\bibitem [{\citenamefont {Zhang}\ \emph {et~al.}(2017)\citenamefont {Zhang},
  \citenamefont {Li}, \citenamefont {Ma}, \citenamefont {Liu}, \citenamefont
  {Cheng}, \citenamefont {Han},\ and\ \citenamefont
  {Zhao}}]{zhang2017efficient}%
  \BibitemOpen
  \bibfield  {author} {\bibinfo {author} {\bibfnamefont {Y.}~\bibnamefont
  {Zhang}}, \bibinfo {author} {\bibfnamefont {P.}~\bibnamefont {Li}}, \bibinfo
  {author} {\bibfnamefont {C.}~\bibnamefont {Ma}}, \bibinfo {author}
  {\bibfnamefont {S.}~\bibnamefont {Liu}}, \bibinfo {author} {\bibfnamefont
  {H.}~\bibnamefont {Cheng}}, \bibinfo {author} {\bibfnamefont
  {L.}~\bibnamefont {Han}},\ and\ \bibinfo {author} {\bibfnamefont
  {J.}~\bibnamefont {Zhao}},\ }\bibfield  {title} {\bibinfo {title} {Efficient
  generation of vector beams by calibrating the phase response of a spatial
  light modulator},\ }\href@noop {} {\bibfield  {journal} {\bibinfo  {journal}
  {Applied optics}\ }\textbf {\bibinfo {volume} {56}},\ \bibinfo {pages} {4956}
  (\bibinfo {year} {2017})}\BibitemShut {NoStop}%
\bibitem [{\citenamefont {Liu}\ \emph {et~al.}(2018)\citenamefont {Liu},
  \citenamefont {Qi}, \citenamefont {Zhang}, \citenamefont {Li}, \citenamefont
  {Wu}, \citenamefont {Han},\ and\ \citenamefont {Zhao}}]{liu2018highly}%
  \BibitemOpen
  \bibfield  {author} {\bibinfo {author} {\bibfnamefont {S.}~\bibnamefont
  {Liu}}, \bibinfo {author} {\bibfnamefont {S.}~\bibnamefont {Qi}}, \bibinfo
  {author} {\bibfnamefont {Y.}~\bibnamefont {Zhang}}, \bibinfo {author}
  {\bibfnamefont {P.}~\bibnamefont {Li}}, \bibinfo {author} {\bibfnamefont
  {D.}~\bibnamefont {Wu}}, \bibinfo {author} {\bibfnamefont {L.}~\bibnamefont
  {Han}},\ and\ \bibinfo {author} {\bibfnamefont {J.}~\bibnamefont {Zhao}},\
  }\bibfield  {title} {\bibinfo {title} {Highly efficient generation of
  arbitrary vector beams with tunable polarization, phase, and amplitude},\
  }\href@noop {} {\bibfield  {journal} {\bibinfo  {journal} {Photonics
  Research}\ }\textbf {\bibinfo {volume} {6}},\ \bibinfo {pages} {228}
  (\bibinfo {year} {2018})}\BibitemShut {NoStop}%
\bibitem [{\citenamefont {Guo}\ \emph {et~al.}(2021)\citenamefont {Guo},
  \citenamefont {Feng}, \citenamefont {Fu},\ and\ \citenamefont
  {Min}}]{guo2021generation}%
  \BibitemOpen
  \bibfield  {author} {\bibinfo {author} {\bibfnamefont {L.}~\bibnamefont
  {Guo}}, \bibinfo {author} {\bibfnamefont {Z.}~\bibnamefont {Feng}}, \bibinfo
  {author} {\bibfnamefont {Y.}~\bibnamefont {Fu}},\ and\ \bibinfo {author}
  {\bibfnamefont {C.}~\bibnamefont {Min}},\ }\bibfield  {title} {\bibinfo
  {title} {Generation of vector beams array with a single spatial light
  modulator},\ }\href@noop {} {\bibfield  {journal} {\bibinfo  {journal}
  {Optics Communications}\ }\textbf {\bibinfo {volume} {490}},\ \bibinfo
  {pages} {126915} (\bibinfo {year} {2021})}\BibitemShut {NoStop}%
\bibitem [{\citenamefont {Chen}\ and\ \citenamefont
  {She}(2010)}]{chen2010single}%
  \BibitemOpen
  \bibfield  {author} {\bibinfo {author} {\bibfnamefont {L.}~\bibnamefont
  {Chen}}\ and\ \bibinfo {author} {\bibfnamefont {W.}~\bibnamefont {She}},\
  }\bibfield  {title} {\bibinfo {title} {Single-photon spin-orbit entanglement
  violating a bell-like inequality},\ }\href@noop {} {\bibfield  {journal}
  {\bibinfo  {journal} {JOSA B}\ }\textbf {\bibinfo {volume} {27}},\ \bibinfo
  {pages} {A7} (\bibinfo {year} {2010})}\BibitemShut {NoStop}%
\bibitem [{\citenamefont {Cardano}\ \emph {et~al.}(2012)\citenamefont
  {Cardano}, \citenamefont {Karimi}, \citenamefont {Slussarenko}, \citenamefont
  {Marrucci}, \citenamefont {de~Lisio},\ and\ \citenamefont
  {Santamato}}]{cardano2012polarization}%
  \BibitemOpen
  \bibfield  {author} {\bibinfo {author} {\bibfnamefont {F.}~\bibnamefont
  {Cardano}}, \bibinfo {author} {\bibfnamefont {E.}~\bibnamefont {Karimi}},
  \bibinfo {author} {\bibfnamefont {S.}~\bibnamefont {Slussarenko}}, \bibinfo
  {author} {\bibfnamefont {L.}~\bibnamefont {Marrucci}}, \bibinfo {author}
  {\bibfnamefont {C.}~\bibnamefont {de~Lisio}},\ and\ \bibinfo {author}
  {\bibfnamefont {E.}~\bibnamefont {Santamato}},\ }\bibfield  {title} {\bibinfo
  {title} {Polarization pattern of vector vortex beams generated by q-plates
  with different topological charges},\ }\href@noop {} {\bibfield  {journal}
  {\bibinfo  {journal} {Applied optics}\ }\textbf {\bibinfo {volume} {51}},\
  \bibinfo {pages} {C1} (\bibinfo {year} {2012})}\BibitemShut {NoStop}%
\bibitem [{\citenamefont {Fu}\ \emph {et~al.}(2016)\citenamefont {Fu},
  \citenamefont {Wang},\ and\ \citenamefont {Gao}}]{fu2016generating}%
  \BibitemOpen
  \bibfield  {author} {\bibinfo {author} {\bibfnamefont {S.}~\bibnamefont
  {Fu}}, \bibinfo {author} {\bibfnamefont {T.}~\bibnamefont {Wang}},\ and\
  \bibinfo {author} {\bibfnamefont {C.}~\bibnamefont {Gao}},\ }\bibfield
  {title} {\bibinfo {title} {Generating perfect polarization vortices through
  encoding liquid-crystal display devices},\ }\href@noop {} {\bibfield
  {journal} {\bibinfo  {journal} {Applied optics}\ }\textbf {\bibinfo {volume}
  {55}},\ \bibinfo {pages} {6501} (\bibinfo {year} {2016})}\BibitemShut
  {NoStop}%
\bibitem [{\citenamefont {Mandal}\ \emph {et~al.}(2020)\citenamefont {Mandal},
  \citenamefont {Maji},\ and\ \citenamefont {Brundavanam}}]{mandal2020common}%
  \BibitemOpen
  \bibfield  {author} {\bibinfo {author} {\bibfnamefont {A.}~\bibnamefont
  {Mandal}}, \bibinfo {author} {\bibfnamefont {S.}~\bibnamefont {Maji}},\ and\
  \bibinfo {author} {\bibfnamefont {M.~M.}\ \bibnamefont {Brundavanam}},\
  }\bibfield  {title} {\bibinfo {title} {Common-path generation of stable
  cylindrical perfect vector vortex beams of arbitrary order},\ }\href@noop {}
  {\bibfield  {journal} {\bibinfo  {journal} {Optics Communications}\ }\textbf
  {\bibinfo {volume} {469}},\ \bibinfo {pages} {125807} (\bibinfo {year}
  {2020})}\BibitemShut {NoStop}%
\bibitem [{\citenamefont {Beijersbergen}\ \emph {et~al.}(1994)\citenamefont
  {Beijersbergen}, \citenamefont {Coerwinkel}, \citenamefont {Kristensen},\
  and\ \citenamefont {Woerdman}}]{beijersbergen1994helical}%
  \BibitemOpen
  \bibfield  {author} {\bibinfo {author} {\bibfnamefont {M.}~\bibnamefont
  {Beijersbergen}}, \bibinfo {author} {\bibfnamefont {R.}~\bibnamefont
  {Coerwinkel}}, \bibinfo {author} {\bibfnamefont {M.}~\bibnamefont
  {Kristensen}},\ and\ \bibinfo {author} {\bibfnamefont {J.}~\bibnamefont
  {Woerdman}},\ }\bibfield  {title} {\bibinfo {title} {Helical-wavefront laser
  beams produced with a spiral phaseplate},\ }\href@noop {} {\bibfield
  {journal} {\bibinfo  {journal} {Optics communications}\ }\textbf {\bibinfo
  {volume} {112}},\ \bibinfo {pages} {321} (\bibinfo {year}
  {1994})}\BibitemShut {NoStop}%
\bibitem [{\citenamefont {Zhu}\ and\ \citenamefont
  {Wang}(2014)}]{zhu2014arbitrary}%
  \BibitemOpen
  \bibfield  {author} {\bibinfo {author} {\bibfnamefont {L.}~\bibnamefont
  {Zhu}}\ and\ \bibinfo {author} {\bibfnamefont {J.}~\bibnamefont {Wang}},\
  }\bibfield  {title} {\bibinfo {title} {Arbitrary manipulation of spatial
  amplitude and phase using phase-only spatial light modulators},\ }\href@noop
  {} {\bibfield  {journal} {\bibinfo  {journal} {Scientific reports}\ }\textbf
  {\bibinfo {volume} {4}},\ \bibinfo {pages} {7441} (\bibinfo {year}
  {2014})}\BibitemShut {NoStop}%
\bibitem [{\citenamefont {Gamel}\ and\ \citenamefont
  {James}(2012)}]{gamel2012measures}%
  \BibitemOpen
  \bibfield  {author} {\bibinfo {author} {\bibfnamefont {O.}~\bibnamefont
  {Gamel}}\ and\ \bibinfo {author} {\bibfnamefont {D.~F.}\ \bibnamefont
  {James}},\ }\bibfield  {title} {\bibinfo {title} {Measures of quantum state
  purity and classical degree of polarization},\ }\href@noop {} {\bibfield
  {journal} {\bibinfo  {journal} {Physical Review A}\ }\textbf {\bibinfo
  {volume} {86}},\ \bibinfo {pages} {033830} (\bibinfo {year}
  {2012})}\BibitemShut {NoStop}%
\bibitem [{\citenamefont {De~Zela}(2014)}]{de2014relationship}%
  \BibitemOpen
  \bibfield  {author} {\bibinfo {author} {\bibfnamefont {F.}~\bibnamefont
  {De~Zela}},\ }\bibfield  {title} {\bibinfo {title} {Relationship between the
  degree of polarization, indistinguishability, and entanglement},\ }\href@noop
  {} {\bibfield  {journal} {\bibinfo  {journal} {Physical Review A}\ }\textbf
  {\bibinfo {volume} {89}},\ \bibinfo {pages} {013845} (\bibinfo {year}
  {2014})}\BibitemShut {NoStop}%
\bibitem [{\citenamefont {Qian}\ and\ \citenamefont
  {Eberly}(2011)}]{qian2011entanglement}%
  \BibitemOpen
  \bibfield  {author} {\bibinfo {author} {\bibfnamefont {X.-F.}\ \bibnamefont
  {Qian}}\ and\ \bibinfo {author} {\bibfnamefont {J.}~\bibnamefont {Eberly}},\
  }\bibfield  {title} {\bibinfo {title} {Entanglement and classical
  polarization states},\ }\href@noop {} {\bibfield  {journal} {\bibinfo
  {journal} {Optics letters}\ }\textbf {\bibinfo {volume} {36}},\ \bibinfo
  {pages} {4110} (\bibinfo {year} {2011})}\BibitemShut {NoStop}%
\bibitem [{\citenamefont {Moreno}\ \emph {et~al.}(2012)\citenamefont {Moreno},
  \citenamefont {Davis}, \citenamefont {Hernandez}, \citenamefont {Cottrell},\
  and\ \citenamefont {Sand}}]{moreno2012complete}%
  \BibitemOpen
  \bibfield  {author} {\bibinfo {author} {\bibfnamefont {I.}~\bibnamefont
  {Moreno}}, \bibinfo {author} {\bibfnamefont {J.~A.}\ \bibnamefont {Davis}},
  \bibinfo {author} {\bibfnamefont {T.~M.}\ \bibnamefont {Hernandez}}, \bibinfo
  {author} {\bibfnamefont {D.~M.}\ \bibnamefont {Cottrell}},\ and\ \bibinfo
  {author} {\bibfnamefont {D.}~\bibnamefont {Sand}},\ }\bibfield  {title}
  {\bibinfo {title} {Complete polarization control of light from a liquid
  crystal spatial light modulator},\ }\href@noop {} {\bibfield  {journal}
  {\bibinfo  {journal} {Optics Express}\ }\textbf {\bibinfo {volume} {20}},\
  \bibinfo {pages} {364} (\bibinfo {year} {2012})}\BibitemShut {NoStop}%
\end{thebibliography}
\end{document}